\documentclass[aps,prd,showpacs]{revtex4}

\usepackage{amsmath,amssymb,amsfonts,amsthm,bbm} 
\usepackage{color}
\usepackage{graphicx}
\usepackage{pstricks,pstricks-add}
\usepackage{pst-plot}
\usepackage[hang,nooneline]{subfigure}
\usepackage{slashbox}

\definecolor{dark-green}{rgb}{0,0.7,0}
\definecolor{dark-blue}{rgb}{0,0.2,0.5}
\definecolor{med-blue}{rgb}{0,0.7,1}
\definecolor{mblue}{rgb}{0,0.2,1}
\definecolor{cnc}{rgb}{0.8,0,0}
\definecolor{light-red}{rgb}{1,0.8,0.8}
\definecolor{dark-yellow}{rgb}{1,0.8,0}
\definecolor{light-blue}{rgb}{0.8,0.9,1}
\definecolor{grey}{rgb}{0.211,0.211,0.211}
\definecolor{verylight-blue}{rgb}{0.93,0.95,1}
\definecolor{light-yellow}{rgb}{1,0.9,0.8}

\newcommand{\vek}[2]{ \left( \begin{smallmatrix} #1\\#2 \end{smallmatrix} \right) }

\begin{document}

\title{Analytic solutions of the geodesic equation in higher dimensional static spherically
symmetric space–times}




\author{Eva Hackmann}
\email{hackmann@zarm.uni-bremen.de}
\altaffiliation{ZARM, Universit\"at Bremen, Am Fallturm, 28359 Bremen, Germany}
\author{Valeria Kagramanova}
\email{kavageo@theorie.physik.uni-oldenburg.de}
\altaffiliation{Institut f\"ur Physik, Universit\"at Oldenburg, 26111 Oldenburg, Germany}
\author{Jutta Kunz}
\email{kunz@theorie.physik.uni-oldenburg.de}
\altaffiliation{Institut f\"ur Physik, Universit\"at Oldenburg, 26111 Oldenburg, Germany}
\author{Claus L{\"a}mmerzahl}
\email{laemmerzahl@zarm.uni-bremen.de}
\altaffiliation{ZARM, Universit\"at Bremen, Am Fallturm, 28359 Bremen, Germany}

\date\today

\begin{abstract}

The complete analytical solutions of the geodesic equation of massive
test particles in higher dimensional Schwarzschild, Schwarzschild--(anti)de
Sitter, Reissner--Nordstr\"om and Reissner--Nordstr\"om--(anti)de Sitter
space--times are presented. Using the Jacobi inversion problem
restricted to the theta divisor the explicit solution is given in
terms of Kleinian sigma functions. The derived orbits depend on
the structure of the roots of the characteristic polynomials which depend on the 
particle's energy and angular momentum, on the mass and the charge of the
gravitational source, and the cosmological constant. We discuss the general structure of the orbits and show that due to the specific dimension--independent form of the angular momentum and the cosmological force a rich variety of orbits can emerge only in four and five dimensions. We present explicit analytical solutions for orbits up to 11 dimensions. A particular feature of Reissner--Nordstr\"om space--times is that bound and escape orbits traverse through different universes. 


\end{abstract}

\pacs{04.20.Jb, 02.30.Hq}

\maketitle

\section{Introduction and Motivation}

The idea of higher dimensions has a long history.
Already in the 1920s Kaluza and Klein unified electromagnetism and gravity 
by introducing a $5$th dimension,
while today string theory, 
as a promising candidate for the quantum theory of gravity
and for the unification of all interactions,
requires higher dimensions for its mathematical consistency.
Following Kaluza and Klein the higher dimensions may be considered small and compact 
and therefore not observable. 
Alternatively, in braneworld scenarios, large higher dimensions
are invoked, where only gravity
is allowed to penetrate the extra dimensions~\cite{AH429-436,RS23-17}, 
while the other fundamental interactions are considered to be trapped on the 3-brane
forming our universe.

In such braneworld models the four-dimensional gravitational coupling $G$,
corresponding to the Planck scale of $10^{19}$ GeV, 
does not set the fundamental scale. Instead, the fundamental
scale is assumed to be on the order of the electroweak scale,
and the extra dimensions then serve to solve the hierarchy problem
by invoking either flat compact extra dimensions
as in the Arkani-Hamed--Dimopoulos--Dvali model \cite{AH429-436}
or warped extra dimensions as in Randall--Sundrum models 
\cite{RS23-17}.
Both type of models predict the production and evaporation 
of mini black holes in high-energy collisions \cite{EG_PRL66_2002} 
at energies in the TeV range accessible at the LHC at CERN
or by observing high--energy cosmic rays in the atmosphere 
\cite{FengShapere_2002, Uehara_2002}.

The study of higher dimensional space-times is also of relevance
from another point of view.
The AdS/CFT correspondence \cite{Maldacena, GuKlePo98} suggests
a duality between a string theory 
involving gravity in Anti-de Sitter (AdS) space,
and a conformal field theory (CFT) living on its boundary.
Study of the phases of higher dimensional AdS black holes, for instance,
can then shed light on the properties of strongly coupled gauge theories.

One of the first papers dedicated to the problem 
of finding solutions of the Einstein equations in higher dimensions 
both in vacuum and coupled to matter fields was by Tangherlini,
who presented the Schwarzschild and Reissner-Nordstr\"om solutions
in $n$ dimensions~\cite{Tang63}. 
Two decades later Myers and Perry~\cite{MyersPerryAnn172} 
generalized the rotating Kerr black hole solutions to higher dimensions. 
The metric of a rotating black hole with a cosmological constant 
was derived in~\cite{Hawking1999AdS} in 5 dimensions and 
in~\cite{GibbonsLuePagePRL93_04} for arbitrary dimensions,
while the general Kerr--NUT--(A)dS solution was
presented in~\cite{ChenLuPopeAllDimKerrNUTAdS}.
A review of black hole solutions in higher dimensional vacuum gravity
was given by Emparan and Reall~\cite{BHinHDEmparanReall}.
In contrast, the higher dimensional
charged rotating black holes of Einstein-Maxwell theory
have not yet been obtained in closed form \cite{Horowitz:2005rs}.
Perturbative Einstein-Maxwell rotating black holes
were studied in
\cite{Aliev:2004ec,Aliev:2005np,Aliev:2006yk,NavarroLerida:2007ez},
while non-perturbative black hole solutions were obtained numerically in
\cite{Kunz:2005nm,Kunz:2006eh} and a finite cosmological constant was included
in \cite{Kunz:2007jq,Brihaye:2007bi} 
(see \cite{Kleihaus:2007kc} for a recent review).

To understand the physical properties of solutions of the gravitational field equations 
it is essential to study the orbits of test particles and light rays in these space--times. 
On the one hand, this is important from an observational point of view, 
since only matter and light are observed and, thus, can give insight 
into the physics of a given gravitational field \cite{Ehlers06}. 
The study of the motion of test particles in gravitational fields
also has significant practical applications. 
On the other hand, this study is also important from a fundamental point of view, 
since the motion of matter and light can be used to classify a given space--time, 
to decode its structure and to highlight its characteristics.

When integrating geodesic equations two fundamental questions arise. These concern the separability of
the Hamilton--Jacobi equation and the possibility to solve the equations of motion analytically.
The first question is related to the symmetries of a given space--time. 
Continuous isometries preserving a metric in curved space--times are generated by Killing vector fields. 
For each Killing vector field there exists a conserved quantity. 
For example, when all metric components do not depend on time and the azimuthal angle, 
energy and angular momentum are conserved during a particle's motion. 
In 1968 Carter~\cite{Carter68} discovered a new separation constant in the Kerr space--time 
and, thus, proved the separability of Hamilton--Jacobi equations. 
This peculiarity of the Kerr metric is associated with hidden symmetries 
associated with the Killing--Yano~\cite{Yano52} tensor. 
Geodesic equations are completely integrable,
and Hamilton--Jacobi, Klein--Gordon and Dirac equations are separable in space--times 
with a metric admitting a conformal Killing--Yano (CKY) tensor. 
A CKY tensor generates a separation constant proportional to a quadratic angular momentum of a particle. 
Hidden symmetries and the separation of variables in higher dimensions are discussed in~\cite{KuKrt07,KrtKuPaVa07,FroKu08,KrFroKu08,PaKuVaKrt}.

The existence of analytical solutions for any kind of problem is not just an academic question. 
In fact, analytical solutions offer systematic applications as well as a frame for tests of the  
accuracy and reliability of numerical integrations. 
In 1931 Hagihara \cite{Hagihara31} first analytically integrated the geodesic equation 
of a test particle motion in Schwarzschild gravitational field. 
This solution is given in terms of the Weierstrass $\wp$--function. 
With the same mathematical tools one can solve the geodesic equation 
in a Reissner--Nordstr\"om spacetime \cite{Chandrasekhar83}. 
The solutions of the geodesic equation in a Kerr and Kerr--Newman space--time 
have also been given analytically (see \cite{Chandrasekhar83} for a survey).
Recently two of us found the complete analytical solution of the geodesic equation 
in Schwarzschild--(anti) de Sitter space--times in $4$ dimensions 
\cite{HackmannLaemmerzahl08_PRL, HackmannLaemmerzahl08_PRD}. 
These calculations are based on the inversion problem of the hyperelliptic Abelian integrals
\cite{4goodoldguys}. The equations of motion could be explicitly solved
\cite{HackmannLaemmerzahl08_PRL,HackmannLaemmerzahl08_PRD} by restricting the problem to the theta divisor.
This procedure makes it possible to obtain a one--parameter solution of the inversion problem. 
The approach was suggested by Enolskii, Pronine and Richter
who applied this method to the problem of the double pendulum \cite{EnolskiiPronineRichter03}.

Our interest here focuses on the motion of test particles in higher dimensional 
spherically symmetric space--times.
In Table~\ref{tabelle} we characterize space--times with respect 
to the possiblity to solve the geodesic equation analytically.
\begin{table}[h!]
\begin{center}
\begin{tabular}{l|*{8}{cccccccc}}
\slashbox[48mm]{Space--time}{Dimension} & 4 & 5 & 6 & 7 & 8 & 9 & 10 & 11 & $d \geq 12$  \\ \hline
Schwarzschild                           & $+$ & $+$ & $\oplus$ & $+$ & $-$ & $\oplus$ & $-$ & $\oplus$ & $-$ \\ \hline
Schwarzschild--de Sitter                & $\oplus$ & + & $-$ & $+$ & $-$ & $\oplus$ & $-$ & $\oplus$ & $-$ \\ \hline
Reissner--Nordstr\"om                   & $+$ & $+$ & $-$ & $\oplus$ & $-$ & $-$ & $-$ & $-$ & $-$ \\ \hline
Reissner--Nordstr\"om--de Sitter        & $\oplus$ & $+$ & $-$ & $\oplus$ & $-$ & $-$ & $-$ & $-$ & $-$ \\ 
\end{tabular}
\end{center}
\caption{Characterization of higher dimensional spherically symmetric space--times with respect to the integrability of the geodesic equation. The + denotes the solvability by elliptic functions, $\oplus$ the solvability in terms of hyperelliptic functions, and a $-$ indicates that it is not analytically solvable
by any known method. 
\label{tabelle}}
\end{table}
Space--times whose geodesic equations can be integrated by elliptic functions are marked by a $+$. 
The underlying polynomials $P(r)$ in these equations can be reduced by a simple substitution $r=\sqrt{x}$
or $r=1/\sqrt{x}$ to polynomials of degree $3$ or $4$. 
Polynomials of $4$th order can be subjected to a subsequent substitution $x=1/u + r_4$ 
which reduces $P_4(x)$ to a cubic expression (here $r_4$ is a zero of $P_4(x)$).
The complete integration of the geodesic equation with an underlying polynomial of order $3$ 
can be performed in terms of elliptic functions.
As an example consider the geodesic equation in the Schwarzschild space--time in $7$ dimensions
\begin{equation}
\label{EG_schw7d}
\left(\frac{dr}{d\varphi} \right)^2=\frac{P_6(r)}{r^2 L^2} \quad \text{where} \quad P_6(r)= (E^2 - 1) r^6 - r^4 L^2 + r^4_{\rm S} r^2+ L^2 r^4_{\rm S} \ ,
\end{equation}
with the conserved quantities $E$ and $L$ and $r_{\rm S}=2M$~\footnote{see Section~\ref{Sec:Geodesic} for
the derivation of the geodesic equation and a detailed discussion}.
Introducing the new variable $x=r^2$ reduces equation~\eqref{EG_schw7d} to a new one with a 
cubic polynomial $P_3(x)$
\begin{equation}
\label{EG_schw7d_p3}
\left(\frac{dx}{d\varphi} \right)^2=\frac{4}{L^2}P_3(x) \quad \text{where} \quad P_3(x)= (E^2 - 1) x^3 - x^2 L^2 + r^4_{\rm S} x + L^2 r^4_{\rm S} \ .
\end{equation}
The solutions of equation~\eqref{EG_schw7d_p3} are given in terms of the Weierstrass $\wp$--function.

The symbol $\oplus$ indicates space--times whose geodesic equations contain polynomials of order 5 
(or polynomials reducible to a polynomial of the 5th degree). 
These equations are hyperelliptic and can be solved by the method suggested 
in~\cite{HackmannLaemmerzahl08_PRL, HackmannLaemmerzahl08_PRD}. 
In these papers the method was applied to the integration of geodesics 
in the Schwarzschild--de Sitter space--time in $4$ dimensions 
and in the Schwarzschild space--time in 6 dimensions. 
Here we study the remaining seven $\oplus$ cases.
The ``$-$'' indicates space--times whose equations of motion cannot be integrated analytically 
by any known methods, since the underlying polynomials are of 7th order or higher.

We begin with a short introduction to the new method of hyperelliptic functions 
used for the integration of the equations of motion 
and apply it first to the Schwarzschild space--time in $9$ and $11$ dimensions 
in Section~\ref{Sec:Schw}. 
Then we continue with the Schwarzschild--(anti-)de Sitter space--time
in both $9$ and $11$ dimensions in Section~\ref{Sec:SchwdS}, 
the Reissner--Nordstr\"om (RN) space--time in 7 dimensions in Section~\ref{Sec:RN} 
and the RN--(anti-)de Sitter space--time in $4$ and $7$ dimensions in Section~\ref{Sec:RNdS}. 
We also include the solution of the geodesic equation 
in the 4--dimensional RN--(anti-)de Sitter space--time, 
which also requires the knowledge of the theory of hyperelliptic functions. 
The resulting orbits are classified in terms of the energy and the angular momentum 
of a test particle, the charge of the gravitational source and the cosmological constant.

\section{The geodesic equation}\label{Sec:Geodesic}

\subsection{The space-times}

We restrict ourselves to static spherically symmetric vacuum solutions of the Einstein equations in an arbitrary number of dimensions $d$
\begin{equation}
\label{RNdDS}
ds^2=  f(r) dt^2 - f(r)^{-1} dr^2 - r^2d\Omega^2_{d-2} \ ,
\end{equation}
with
\begin{equation}
\label{f(r)}
f(r)=1-\left(\frac{r_{\rm S}}{r}\right)^{d-3}-\frac{2\Lambda r^2}{(d-1)(d-2)} + \left(\frac{q}{r}\right)^{2(d-3)}  \ ,
\end{equation}
where $M$ and $q$ are the mass and charge, respectively, of a gravitational source, $\Lambda$ is a cosmological constant, $r_{\rm S}=2 M$, and $d\Omega^2_1=d\varphi^2$ and $d\Omega^2_{i+1}=d\theta^2_i+\sin^2\theta_id\Omega^2_i$ for $i\ge1$. These solutions are uniquely characterized by their mass, charge and cosmological constant \cite{GibbonsIdaShiromizu02}.

\subsection{The geodesic equation}

The motion of a particle is given by the geodesic equation 
\begin{equation}
D_v v = 0 \label{EOM} \ ,
\end{equation}
where $v$ is the 4--velocity of the particle and $D$ is the covariant derivative based on the space--time metric. We also use the normalization $g(v, v) = \epsilon$ where $\epsilon = 1$ for massive particles and $\epsilon=0$ for light. 

Due to spherical symmetry of the space--time the motion of a test particle can be restricted to the equatorial plane defined by $\theta_i = \pi/2$, $i \geq 1$. There are two conserved quantities: the dimensionless energy $E$, and the angular momentum $L$ with the dimension of length (both are normalised to the mass of the test particle)
\begin{equation}
E= f(r) \frac{dt}{ds} \ , \qquad L = r^2\frac{d\varphi}{ds} \ .
\end{equation}
Equation~\eqref{EOM} combines the equations describing the dynamics of the particle's motion
\begin{eqnarray}
\left(\frac{dr}{d\varphi}\right)^2 &=& \frac{r^4}{L^2} \left(E^2 - \left(1 - f(r)\right) \left(\epsilon+\frac{L^2}{r^2} \right) \right) \ , \label{EOMrphi} \\
\left(\frac{dr}{ds}\right)^2 & = & E^2 - f(r) \left(\epsilon+\frac{L^2}{r^2} \right) \label{drds} \\ 
\left(\frac{dr}{dt}\right)^2 & = & \frac{ f^2(r)}{E^2} \left(E^2 - f(r) \left(\epsilon+\frac{L^2}{r^2} \right)\right) \, . \label{drdt}
\end{eqnarray}
We also introduce the effective potential $V_{\rm eff}$ which can be read off~\eqref{drds}
\begin{equation}
\label{Veff}
V_{\rm eff}= \left(1 - \left(\frac{r_{\rm S}}{r}\right)^{d-3} -\frac{2\Lambda r^2}{(d-1)(d-2)} + \left(\frac{q}{r}\right)^{2(d-3)} \right) \left(\epsilon+\frac{L^2}{r^2} \right) \ .
\end{equation}
It is obvious that at any horizon, which is defined by $f(r) = 0$, the effective potential vanishes. 

For convenience we rewrite~\eqref{EOMrphi} with new dimensionless parameters and a dimensionless coordinate $\tilde{r}$
\begin{equation}
\lambda=\frac{r^2_{\rm S}}{L^2} \, , \quad \mu = E^2 \, , \quad \eta=\frac{q^2}{r_{\rm S}^2} \, , \quad \tilde\Lambda = \Lambda r_{\rm S}^2 \, , \qquad \tilde{r} = \frac{r}{r_{\rm S}} \,
\end{equation}
and obtain
\begin{equation}
\left(\frac{d\tilde{r}}{d\varphi}\right)^2 = \lambda \tilde{r}^4 \left[\mu - \left(1 - \frac{1}{\tilde{r}^{d-3}} - \frac{2 \tilde\Lambda\tilde{r}^2}{(d-1)(d-2)} + \frac{\eta^{d-3}}{\tilde{r}^{2(d-3)}}\right) \left(\epsilon + \frac{1}{\lambda\tilde{r}^2} \right) \right] \, . \label{EOMrphinorm}
\end{equation}
The right hand side of this equation can be written in the form $P_n(\tilde{r})/\tilde{r}^m$, where $P_n$ is a polynomial of order $n$, see Table \ref{tabellePn}. 
\begin{table}[h!]
\begin{center}
\begin{tabular}{r|cc}
& $\eta = 0$ & $\eta \neq 0$ \\ \hline
$\tilde\Lambda = 0$ & $\left(\frac{d\tilde{r}}{d\varphi}\right)^2 = \frac{P_{d-1}}{\tilde{r}^{d-5}}$ & $\left(\frac{d\tilde{r}}{d\varphi}\right)^2 = \frac{P_{2(d-2)}}{\tilde{r}^{2(d-4)}}$ \\
$\tilde\Lambda \neq 0$ & $\left(\frac{d\tilde{r}}{d\varphi}\right)^2 = \frac{P_{d+1}}{\tilde{r}^{d-5}}$ & $\left(\frac{d\tilde{r}}{d\varphi}\right)^2 = \frac{P_{2(d-1)}}{\tilde{r}^{2(d-4)}}$ 
\end{tabular}
\end{center}
\caption{Order of the polynomials appearing in the equation of motion~\eqref{EOMrphinorm} in various $d$--dimensional space-times. \label{tabellePn}}
\end{table}

Depending on the space--time under consideration the equation of motion~\eqref{EOMrphinorm} after some transformation $\tilde{r}=\chi(x)$ can be reduced to one of the two forms
      \begin{subequations}\label{rP_5}
      \begin{eqnarray}
      \left(x\frac{dx}{d\varphi}\right)^2=P_5(x) \label{rP_5a} \\
      \left(\frac{dx}{d\varphi}\right)^2=P_5(x) \label{rP_5b} \ ,
      \end{eqnarray}
      \end{subequations}
where $P_5(x)=a_5 x^5 + a_4 x^4 + a_3 x^3 + a_2 x^2 + a_1 x + a_0$. 

Because of \eqref{EOMrphinorm} space--times with odd number of dimensions give polynomials $P_n$ containing even powers of $\tilde r$ only. Therefore we simplify \eqref{EOMrphinorm} by introducing $u = 1/\tilde{r}^2$ what reduces the degree of $P_{n}$ by a factor of 2 
\begin{eqnarray}
\frac{1}{4}\left(\frac{du}{d\varphi}\right)^2 & = & - \eta^{d-3}u^{d-1} - \epsilon \lambda \eta^{d-3}u^{d-2} + u^{\frac{1}{2}(d+1)} + \epsilon \lambda u^{\frac{1}{2}(d-1)} -u^2 \nonumber  \\ 
 & & + u \left(\lambda(\mu - \epsilon) + \frac{2 \tilde\Lambda}{(d-1)(d-2)}\right) + \frac{2 \epsilon \lambda \tilde\Lambda}{(d-1)(d-2)} \, . \label{EOMrphinormu}
\end{eqnarray} 
The number of positive real zeros of the polynomials on the right hand side of \eqref{EOMrphinormu} and \eqref{EOMrphinorm} coincide. 
Space--times with an even number of dimensions, as e.g. 4--dimensional Reissner--Nordstr\"om--de Sitter space--time considered in Section~\ref{Sec:RNdS4D}, contain polynomials of odd degree so that this substitution is not useful.  

\subsection{The topology of the orbits}

Owing to the square on the left--hand--side of \eqref{drds} a test particle is only  allowed to move in the region restricted by the condition $E^2 \geq V_{\rm eff}$. Therefore, the motion is essentially characterized by the number of real positive zeros the polynomials $P_n(\tilde{r})$ possess. The number of real positive roots of a polynomial $P_n(\tilde{r})$ can easily be determined by using Descartes' rule of signs. 

If $P_n(\tilde{r})$ possesses no real positive zero and $P_n(\tilde{r}) > 0$, then the particle is coming from infinity and moves directly to the origin. Such an orbit we call a {\it terminating escape orbit}. For $P_n(\tilde{r}) < 0$ no motion is possible. One real positive zero means either that the particle starts at a finite coordinate distance and ends at $\tilde{r} = 0$, what we call a {\it terminating bound} orbit, or it moves on an {\it escape  orbit} with a finite impact parameter. For two real positive zeros we may have two cases: (i) if the polynomial is positive between the two zeros we have a {\it periodic bound orbit} like a planetary orbit, (ii) if the polynomial is negative between the two zeros we have an escape orbit or a terminating bound orbit. The structure of the orbits for more than two positive zeros can be obtained analogously. The actual orbit depends on the chosen initial position. 

Below we discuss in more detail the structure of the orbits for the various dimensions and the allowed values of the energy, angular momentum and cosmological constant. As one result we see that for all considered space--times the richest structure of orbits can be found in four and five dimensions. The reason for that is that only in these lower dimensions various terms combine such that more possibilities arise to obtain different signs for coefficients of the polynomial. For higher dimensions all terms disentangle and the structure of the orbits becomes very simple and is the same for all dimensions. That means that it is the particular structure of the angular momentum part and the cosmological part both being independent of the space--time dimensions, which single out the lower dimensions.

\subsubsection{Schwarzschild space--times}

For the geodesics in Schwarzschild space-times in arbitrary dimensions we have
\begin{equation}
\left(\frac{d\tilde{r}}{d\varphi}\right)^2 = \frac{1}{\tilde{r}^{d-5}} \left(\lambda(\mu - 1) \tilde{r}^{d-1} - \tilde{r}^{d-3} + \lambda \tilde{r}^2 + 1\right) \ , 
\end{equation}
which yields the types of orbits listed in Table~\ref{TablSchw}.

\begin{table}[h!]
\begin{center}
\begin{tabular}{cccl}
dimension & polynomial & parameters & orbits \\ \hline
$d = 4$ & $\lambda(\mu - 1) \tilde{r}^3 + \lambda \tilde{r}^2 - \tilde{r} + 1$ & $\mu < 1$ & periodic bound, terminating bound \\
& & $\mu > 1$ & escape, terminating bound, terminating escape \\ \hline
$d = 5$ & $\lambda(\mu - 1) \tilde{r}^4 + (\lambda - 1) \tilde{r}^2 + 1$ & $\mu < 1$,  $\lambda < 1$ & terminating bound \\
& & $\mu <1$, $\lambda > 1$ & terminating bound  \\
 & & $\mu > 1$, $\lambda < 1$ & terminating bound, escape, terminating escape \\
& & $\mu >1$, $\lambda > 1$ & terminating escape \\ \hline
$d \geq 6$ & $\lambda(\mu - 1) \tilde{r}^{d-1} - \tilde{r}^{d-3} + \lambda \tilde{r}^2 + 1$ & $\mu < 1$ & terminating bound \\
& & $\mu > 1$ & terminating bound, escape, terminating escape \\ \hline
\end{tabular} 
\end{center}
\caption{Types of geodesics in Schwarzschild space--time \label{TablSchw}}
\end{table}

Only in four dimensions there are periodic bound orbits. There are no stable configurations in higher dimensions. 
\subsubsection{Schwarzschild--de Sitter space--times}

The geodesic equation in Schwarzschild--de Sitter space--time in higher dimensions
\begin{equation}
\left(\frac{d\tilde{r}}{d\varphi}\right)^2 = \frac{1}{\tilde{r}^{d-5}} \left[\frac{2 \lambda \tilde\Lambda}{(d-1)(d-2)} \tilde{r}^{d+1} + \left(\lambda (\mu - 1) + \frac{2 \tilde\Lambda}{(d-1)(d-2)}\right) \tilde{r}^{d-1} - \tilde{r}^{d-3} + \lambda \tilde{r}^{2} + 1\right] \label{GeodesicSchwarzschilddeSitter}
\end{equation}
contains an additional term $\tilde\Lambda \tilde{r}^{d+1}$ as compared to the Schwarzschild case. For large $\tilde{r}$ this term is dominant. For small negative $\Lambda$ all escape orbits turn into periodic bound orbits, and for small positive $\Lambda$ some periodic bound orbits (slightly below $\mu = 1$) will turn into escape orbits. A huge positive $\Lambda$ might surpass the angular momentum barrier. We skip the detailed discussion of all subcases and just remark that only for $d = 4$ and $5$ different shapes of orbits occur due to the mixing of different terms. Such mixing no longer occurs for $d \geq 6$.

\subsubsection{Reissner--Nordstr\"om space--times}

The geodesics in RN space--times
\begin{equation}
\left(\frac{d\tilde{r}}{d\varphi}\right)^2 = \frac{\tilde{r}^4}{\tilde{r}^{2(d-2)}} \left(\lambda (\mu - 1) \tilde{r}^{2(d-2)} - \tilde{r}^{2(d-3)} + \lambda \tilde{r}^{d - 1} + \tilde{r}^{d - 3} - \lambda \eta^{d-3} \tilde{r}^{2} - \eta^{d-3}\right) 
\end{equation}
do not possess any terminating escape orbit. The reason is the charge which leads to gravitational repulsion near the singularity at the origin. Accordingly, we have either periodic bound or escape orbits. A particular feature is that the 4 dimensional RN space--time is the only space--time which allows 2 different periodic bound orbits (see Table~\ref{TablRN}). Both periodic bound orbits are characterized by the same period which is a consequence of the fact that the solution, owing to the order of the polynomial, is given in terms of the Weierstrass $\wp$ function. 

\begin{table}[h!]
\begin{center}
\begin{tabular}{cccl}
dimension & polynomial & parameters & orbits \\ \hline
$d = 4$ & $\lambda (\mu - 1) \tilde{r}^4 + \lambda \tilde{r}^3 - (1 + \lambda \eta) \tilde{r}^{2} + \tilde{r} - \eta$ & $\mu < 1$ & 2 periodic bound \\
& & $1 < \mu$ & periodic bound, escape \\ \hline
$d = 5$ & $\lambda (\mu - 1) \tilde{r}^6  + (\lambda - 1) \tilde{r}^4 + (1 - \lambda \eta^2) \tilde{r}^{2} - \eta^2$ & $\mu < 1$,  $\lambda < 1$, $\lambda \eta^2 > 1$ & not allowed \\
& & $\mu < 1$, $\lambda < 1$, $\lambda \eta^2 < 1$ & periodic bound \\
& & $\mu < 1$, $\lambda > 1$, $\lambda \eta^2 > 1$ & periodic bound \\
& & $\mu < 1$, $\lambda > 1$, $\lambda \eta^2 < 1$ & periodic bound \\
& & $\mu > 1$, $\lambda < 1$, $\lambda \eta^2 > 1$ & escape \\
& & $\mu > 1$, $\lambda < 1$, $\lambda \eta^2 < 1$ & periodic bound, escape \\
& & $\mu > 1$, $\lambda > 1$, $\lambda \eta^2 > 1$ & escape \\ 
& & $\mu > 1$, $\lambda > 1$, $\lambda \eta^2 < 1$ & escape \\ \hline
$d \geq 6$ & $\lambda (\mu - 1) \tilde{r}^{2(d-2)} - \tilde{r}^{2(d-3)} + \lambda \tilde{r}^{d - 1}$ & $\mu < 1$ & periodic bound \\
& $ + \tilde{r}^{d - 3} - \lambda \eta^{d-3} \tilde{r}^{2} - \eta^{d-3}$ & $1 < \mu$ & periodic bound, escape \\ \hline
\end{tabular}
\end{center}
\caption{Types of geodesics in Reissner--Nordstr\"om space--time \label{TablRN}}
\end{table}

\subsubsection{Reissner--Nordstr\"om--de Sitter space--times}

For the addition of a cosmological constant in the geodesic equation in a RN space--time
\begin{equation}
\left(\frac{d\tilde{r}}{d\varphi}\right)^2 = \lambda \tilde{r}^4 \left[\mu - \left(1 - \frac{1}{\tilde{r}^{d-3}} - \frac{2 \tilde\Lambda\tilde{r}^2}{(d-1)(d-2)} + \frac{\eta^{d-3}}{\tilde{r}^{2(d-3)}}\right) \left(\epsilon + \frac{1}{\lambda\tilde{r}^2} \right) \right] \, . 
\end{equation}
the remarks made after \eqref{GeodesicSchwarzschilddeSitter} also apply. A point to add is that the periods of the two periodic bound orbits appearing in RN space--time now become different.

\subsection{Integration}

Separation of variables in~\eqref{rP_5} yields the hyperelliptic integral
\begin{subequations}\label{varphiP_5}
\begin{equation}
\varphi-\varphi_{\rm in} = \int^x_{x_{\rm in}} \frac{x^\prime dx^\prime}{\sqrt{P_5(x^\prime)}} \label{varphiP_5a} 
\end{equation}
or
\begin{equation}
\varphi-\varphi_{\rm in} = \int^x_{x_{\rm in}} \frac{dx^\prime}{\sqrt{P_5(x^\prime)}} \label{varphiP_5b}
\end{equation}
\end{subequations}
for the physical angle $\varphi$ with $x_{\rm in} = x(\varphi_{\rm in})$ (the pair $(x_{\rm in}, \varphi_{\rm in})$ is an initial point of a test particle motion). This approach allows us to obtain an explicit solution of the inversion problem by restricting the problem to the theta--divisor (zeros of the theta function). As a result we obtain the complete set of analytically given orbits of test particles. 

There are two important points: The first is the ambiguous definition of the integrand because of two branches of the square root. The second is the periodicity of $x(\varphi)$ following from the requirement of $x(\varphi)$ to be independent of the integration path.
That means that for a closed integration path $\gamma$ and
$\omega=\oint_{\gamma} \frac{x^\prime dx^\prime}{\sqrt{P_5(x^\prime)}}$ or $\omega=\oint_{\gamma} \frac{dx^\prime}{\sqrt{P_5(x^\prime)}}$, respectively, the function $x(\varphi)$ has to fulfill
\begin{equation}
\label{varphiomega}
\varphi-\varphi_{\rm in} = \int^x_{x_{\rm in}} \frac{x^\prime dx^\prime}{\sqrt{P_5(x^\prime)}} + \omega \,, \quad \text{ or } \quad \varphi-\varphi_{\rm in} = \int^x_{x_{\rm in}} \frac{dx^\prime}{\sqrt{P_5(x^\prime)}} + \omega \,,
\end{equation}
and, thus, has to be a periodic function with period $\omega$.

Owing to some modifications compared to \cite{HackmannLaemmerzahl08_PRD} we shortly review the procedure of explicit integration. 

\section{Solving hyperelliptic integrals}\label{Sec:Solution}

We consider~\eqref{varphiP_5} on the Riemannian surface $y^2=P_5(x)$ of genus $g=2$ and introduce a basis of canonical holomorphic and meromorphic differentials $dz_i$ and $du_i$, respectively,
\begin{align}
\label{diffls}
dz_1 & =  \frac{dx}{\sqrt{P_5(x)}} \ , \qquad & dz_2 & =  \frac{x dx}{\sqrt{P_5(x)}} \ , \\
du_1 & =  \frac{a_3 x + 2 a_4 x^2 + 3 a_5 x^3}{4 \sqrt{P_5(x)}}dx \ , \qquad & du_2 & =  \frac{x^2 dx}{4 \sqrt{P_5(x)}} \ ,
\end{align}
and real $2 \omega_{ij}, 2 \eta_{ij}$ and imaginary $2 \omega^\prime_{ij}, 2 \eta^\prime_{ij}$ period matrices
\begin{align}
\label{perds}
2 \omega_{ij} & = \oint_{a_j} dz_i \ , \qquad & 2 \omega^\prime_{ij} & = \oint_{b_j} dz_i \ , \\
2 \eta_{ij} & = \oint_{a_j} du_i \ , \qquad & 2 \eta^\prime_{ij} & = \oint_{b_j} du_i \,
\end{align}
with $i, j = 1,2$ and the canonical basis of cycles $(a_1, a_2; b_1, b_2)$. These 4 closed paths defined on a Riemannian surface correspond to periods of a function defined on it (in our case, to the solution of~\eqref{varphiP_5}).
We also need the normalized holomorphic differentials
\begin{equation}
\label{normdiffls}
d\vec{v} = (2 \omega)^{-1} d\vec{z} \, , \qquad d\vec z = \begin{pmatrix} dz_1 \\ dz_2 \end{pmatrix} \, .
\end{equation}
The period matrix of these differentials is given by $(1_2,\tau)$, where $\tau$ is a Riemann matrix defined by $\tau=\omega^{-1} \omega'$.
It is this $\tau$ which contains all the information about the polynomial $P_5$. 

The solutions of \eqref{varphiP_5} can be inferred from the solutions of the Jacobi inversion problem ~\cite{4goodoldguys} which solves
\begin{align}\label{Jacobi}
\varphi_1  & = \int_{x_0}^{x_1} \frac{dx}{\sqrt{P_5(x)}} + \int_{x_0}^{x_2} \frac{dx}{\sqrt{P_5(x)}} \,, \nonumber \\
\varphi_2  & = \int_{x_0}^{x_1} \frac{x dx}{\sqrt{P_5(x)}} + \int_{x_0}^{x_2} \frac{x dx}{\sqrt{P_5(x)}} \,
\end{align}
for $\vec x = (x_1, x_2)^t$ in terms of $\vec\varphi = (\varphi_1, \varphi_2)^t$. This solution is given by 
\begin{equation}\label{Jacobisol}
\begin{split}
x_1+x_2 = & \frac{4}{a_{5}} \wp_{22}(\vec \varphi)\,, \\
x_1 \, x_2 = & - \frac{4}{a_{5}} \wp_{12}(\vec \varphi) \ ,
\end{split}
\end{equation}
where 
\begin{equation}\label{Weier}
\wp_{ij}(\vec z) = - \frac{\partial}{\partial z_i}  \frac{\partial}{\partial z_j} \log \sigma(\vec z) = \frac{\sigma_i(\vec z) \sigma_j(\vec z) - \sigma(\vec z) \sigma_{ij}(\vec z)}{\sigma^2(\vec z)} 
\end{equation}
are the generalized Weierstrass functions ($\sigma_i = \partial\sigma/\partial z_i$), where 
\begin{equation}\label{sigma}
\sigma(\vec z) = C e^{- \frac{1}{2} \vec z^t \eta \omega^{-1} \vec z} \vartheta \left( (2 \omega)^{-1}\vec z + \tau \vec g + \vec h ; \tau \right)\,,
\end{equation}
(the constant $C$ can be given explicitly, see \cite{BEL97}) is the Kleinian $\sigma$ function based on the theta function
\begin{equation}
\vartheta(\vec z;\tau) = \sum_{\vec m \in {\mathbb{Z}}^2} e^{i \pi \vec m^t (\tau \vec m + 2 \vec z)} \, .
\end{equation}
It is the $\vartheta$--function where the matrix $\tau$ comes in. Note that the problem \eqref{Jacobi} contains the problem \eqref{varphiP_5} for $x_2 \to x_0$ and a restriction such that one of the two equations \eqref{Jacobi} can be omitted.

Without restriction we can choose $x_0 = \infty$ (this only induces a reformulation of $\varphi$) and consider the limit $x_2 \to \infty$ to obtain
\begin{equation}
x_1  = \lim_{x_2 \to \infty} \frac{x_1 x_2}{x_1 + x_2}
 = \frac{\sigma(\vec\varphi_\infty) \sigma_{12}(\vec\varphi_\infty) - \sigma_1(\vec\varphi_\infty) \sigma_2(\vec\varphi_\infty)}{\sigma_2^2(\vec\varphi_\infty) - \sigma(\vec\varphi_\infty) \sigma_{22}(\vec\varphi_\infty)} \ , \label{r_1}
\end{equation}
where $\vec\varphi_\infty  = \lim_{x_2 \to \infty} \vec\varphi \overset{\eqref{Jacobi}}{=} \int^{x_1}_{\infty} d\vec z = \int_{x_{\rm in}}^{x_1} d\vec z - \int_{x_{\rm in}}^{\infty} d\vec z$, where $x_{\rm in}$ is an initial point of a test particle motion.

The next step is to restrict Jacobi's inversion problem to a theta--divisor $\Theta_{K_{x_0}}$, the one-dimensional set of zeros of the theta function $\vartheta \left( (2 \omega)^{-1}\vec z + \vec K_{x_0}; \tau \right)$, where $\vec K_{x_0}$ is the Riemann vector associated with the base point $x_0$~\cite{BEL97}. One can check that $(2\omega)^{-1} \varphi_\infty$ is an element of the theta-divisor $\Theta_{\vec{K}_\infty}$ with $\vec{K}_\infty = \tau \vek{1/2}{1/2} + \vek{0}{1/2}$~\cite{Mum, BEL97} and $\sigma(\vec{\varphi}_\infty)=0$. This allows us to write \eqref{r_1} in a simpler form and to express one of the components of $\varphi_\infty$ in terms of the other component.

Now we can write the analytic solution of the geodesic equation as
\begin{equation}
\tilde{r}(\varphi) = \chi(x(\varphi)) = \chi(x_1) = \chi\left( - \frac{\sigma_1(\vec \varphi_\infty)}{\sigma_2(\vec \varphi_\infty)} \right)\,. \label{gensol}
\end{equation}
Eqn.~\eqref{gensol} is an explicit analytical solution of the geodesic equation~\eqref{EOMrphinorm} which we will explicitly work out for the Schwarzschild, Schwarzschild--de Sitter, Reissner--Nordstr\"om and Reissner--Nordstr\"om--de Sitter space--times in higher dimensions.

It is important to note that, depending on the type of the problem \eqref{varphiP_5}, there are two possible {\it physical angles} $\varphi$:
\begin{subequations}\label{varphiP_5_phys}
\begin{eqnarray}
\varphi_{z_2} &=& \int^{x_1}_{x_{in}} dz_2 + \varphi_{\rm in}  \label{varphiP_5_physa} \ , \\
\varphi_{z_1} &=& \int^{x_1}_{x_{in}} dz_1 + \varphi_{\rm in} \label{varphiP_5_physb} \ .
\end{eqnarray}
\end{subequations}
This means that depending on the problem $\vec{\varphi}_{\infty}$, which is an element of the theta-divisor and, therefore, one-dimensional, will take a corresponding form
\begin{subequations}\label{varphi_inf}
\begin{eqnarray}
      \vec \varphi_{\infty, z_2} & =& \begin{pmatrix} \int_{x_{\rm in}}^{x_1} dz_1 - \int_{x_{\rm in}}^{\infty} dz_1 \\ \varphi_{z_2} - \varphi^\prime_{{\rm in}, z_2}  \end{pmatrix}  \label{varphi_infa} \ , \\
      \vec \varphi_{\infty, z_1} & =& \begin{pmatrix} \varphi_{z_1} - \varphi^\prime_{{\rm in}, z_1}  \\ \int_{x_{\rm in}}^{x_1} dz_2 - \int_{x_{\rm in}}^{\infty} dz_2 \end{pmatrix}  \label{varphi_infb} \ ,
\end{eqnarray}
\end{subequations}
where $\varphi^\prime_{{\rm in}, z_2} = \varphi_{\rm in} + \int^\infty_{x_{\rm in}} d z_2$ and $\varphi^\prime_{{\rm in}, z_1} = \varphi_{\rm in} + \int^\infty_{x_{\rm in}} d z_1$.
Note that in $\varphi_{\infty,z_2}$ the first component is a function of the second component and in $\varphi_{\infty,z_1}$ the second component is a function of the first component.

Now we are going to explicitly work out the solutions for various higher dimensional space--times.

\section{Geodesics in higher dimensional Schwarzschild space--times}\label{Sec:Schw}

In this section we present and discuss the possible types of orbits in the space--time of a gravitating source described by the higher dimensional Schwarzschild metric given by \eqref{RNdDS} for $\Lambda = 0$ and $q = 0$.  
We first address the effective potential $V_{\rm eff}$ in these space--times 
and describe the classification of zeros of the polynomial $P_{n}(\tilde{r})$ in~\eqref{EOMrphinorm} in terms of the parameters $\mu$ and $\lambda$. We then present the explicit analytical solution of the geodesic equation for $\epsilon=1$.

\subsection{Schwarzschild space--time in 9 dimensions}\label{Sec:Schw_9D}

\begin{figure}[t]
\begin{center}
\subfigure[][$(\lambda,\mu)$--plot]{\label{sub:lamuS9d}%
\includegraphics[width=0.3\textwidth]{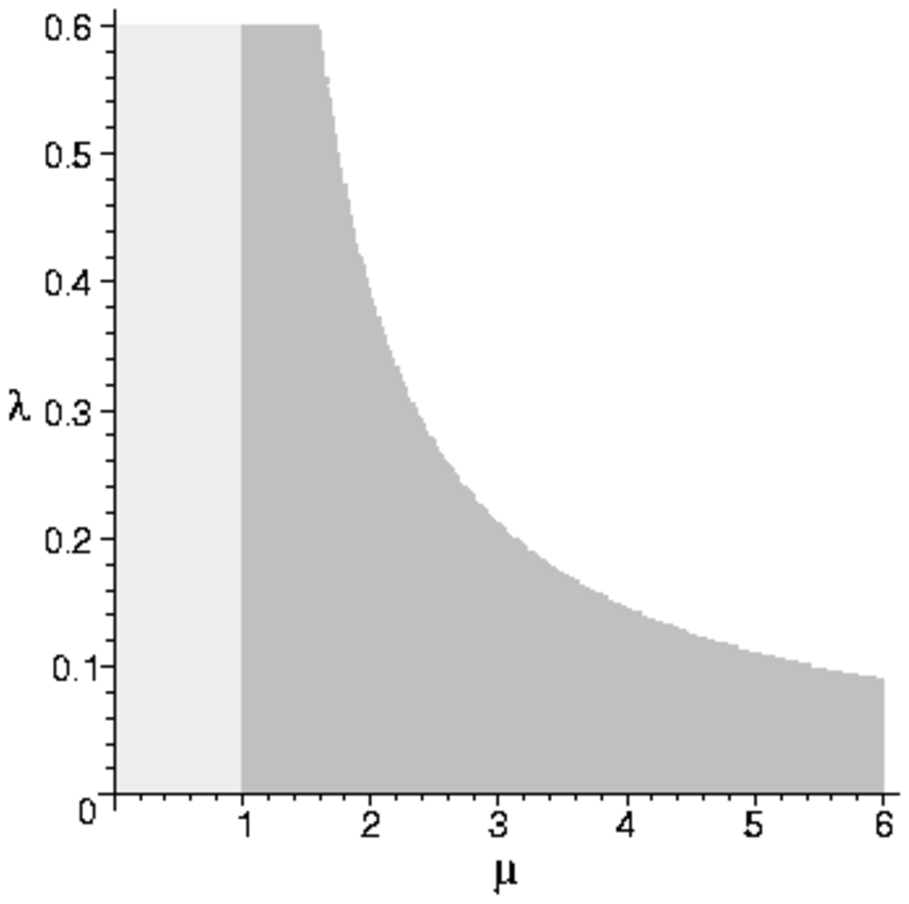}
} \quad %
\subfigure[][Potential for
$\lambda=0.15$]{\label{sub:potS9d}%
\includegraphics[width=0.25\textwidth]{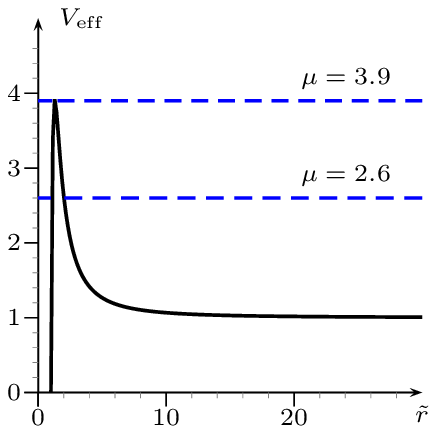}}
\end{center}
\caption{9--dimensional Schwarzschild space-time. \subref{sub:lamuS9d} The zeros of $P_5$ in a $(\lambda,\mu)$--diagram (the number of zeros is encoded in the gray scales: white = 0, light gray = 1, gray = 2 zeros). \subref{sub:potS9d} The effective potential. For the corresponding orbits see Fig.~\ref{orbd9}. \label{lamuSchw9D}}
\end{figure}

The equation of motion~\eqref{EOMrphinormu} for Schwarzschild space--time in 9 dimensions reduces to
\begin{equation}
\label{geodd9}
\left(\frac{d u}{d\varphi}\right)^2=4 u \left(u^4 + \lambda u^3 - u + \lambda(\mu-1)\right) = 4 P_5(u) \ .
\end{equation}
The $\lambda-\mu$ diagram in Fig.~\ref{sub:lamuS9d} shows the number of positive zeros of the polynomial $P_5(u)$ (or, equivalently, $P_8(\tilde{r})$), depending on the parameters $\lambda$ and $\mu$. Here the gray scale code is as follows: gray denotes two, light gray one, and white no positive real zero of $P_5(u)$.

The effective potential~\eqref{Veff} is
\begin{equation}
\label{Veff9S}
V_{\rm eff}= \left(1 - \frac{1}{\tilde{r}^6}\right) \left(1 + \frac{1}{\lambda\tilde{r}^2} \right) \,
\end{equation}

With $z = \tilde{r}^2$ the extrema of the effective potential are given by $0 = z^3 - 3 \lambda z - 4$. The discriminant of this cubic equation is $D = 4 - \lambda^3$. For $D < 0$ there is always one positive real zero given by $z = 2 \sqrt{\lambda} \cos\left(\frac{1}{3} \arccos\frac{2}{\sqrt{\lambda^3}}\right)$. For $D = 0$ the positive zero is given by $z = 4^{2/3}$ and $V_{\rm eff}(4^{1/3})=\frac{75}{64}$. For $D > 0$ we have $z = \left(2 + \sqrt{4 - \lambda^3}\right)^{1/3} + \left(2 - \sqrt{4 - \lambda^3}\right)^{1/3}$.  
For any value of $\lambda$ the second derivative of the potential at the corresponding value of $z$ is negative which implies that the potential has a maximum. There are no stable circular or periodic bound orbits.

Eq.~\eqref{geodd9} is of type~\eqref{rP_5b} and the solution~\eqref{gensol} becomes
\begin{equation}
\tilde{r}(\varphi) =  \sqrt{- \frac{\sigma_2(\vec \varphi_{\infty, z_1})}{\sigma_1(\vec \varphi_{\infty, z_1})} } \ ,  \label{solSchw9D}
\end{equation}
with the angular coordinate $\varphi_{z_1}$ of a test particle from~\eqref{varphiP_5_physb} and $\vec\varphi_{\infty, z_1}$ from equation~\eqref{varphi_infb}; $x_{\rm in}$ here is the starting point of the motion usually chosen to be a zero of $P_5(x)$.

\begin{figure}[th!]
\begin{center}
\subfigure[][$\mu=2.6$: terminating bound orbit and escape orbit]{\includegraphics[width=0.45\textwidth]{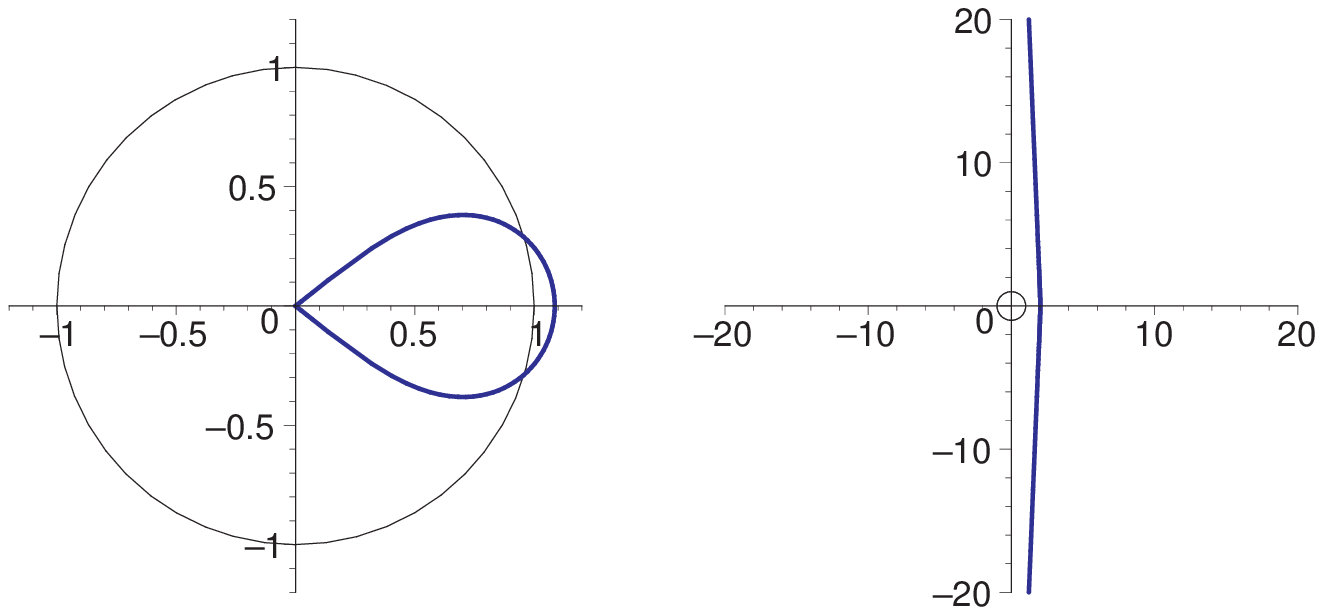}}
\subfigure[][$\mu=3.9$: terminating bound orbit and escape orbit]{\includegraphics[width=0.45\textwidth]{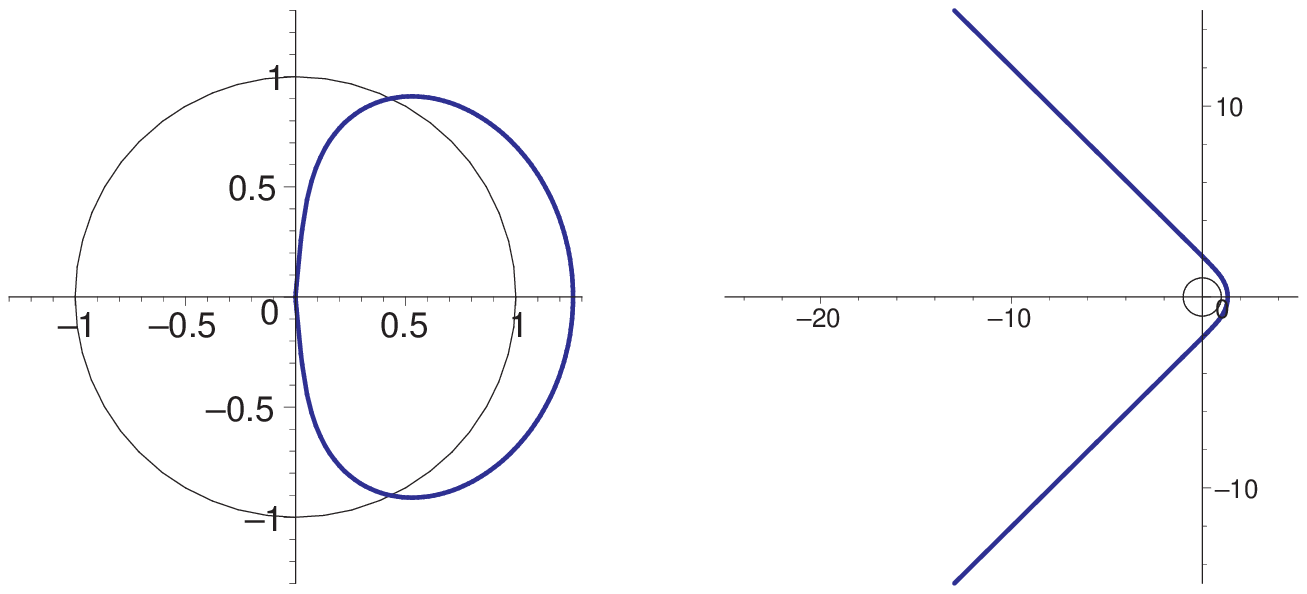}} \\
\subfigure[][$\mu=3.9203$: terminating bound orbit and escape orbit]{\includegraphics[width=0.45\textwidth]{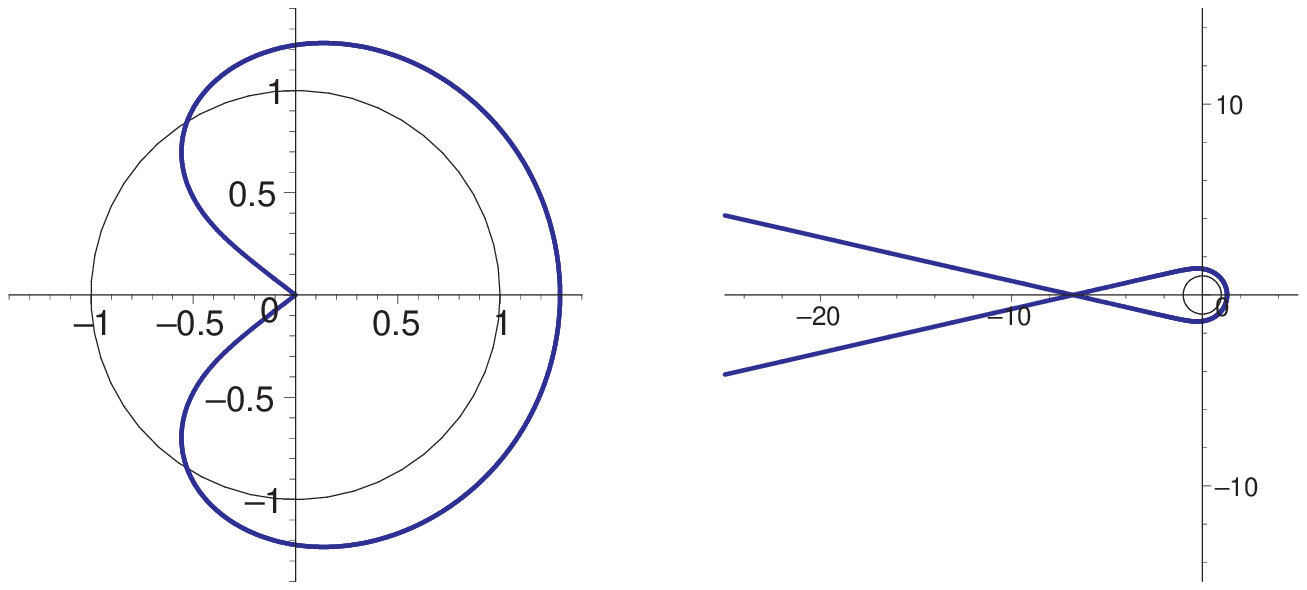}}
\subfigure[][$\mu=3.92046627$: terminating bound orbit and escape orbit]{\includegraphics[width=0.45\textwidth]{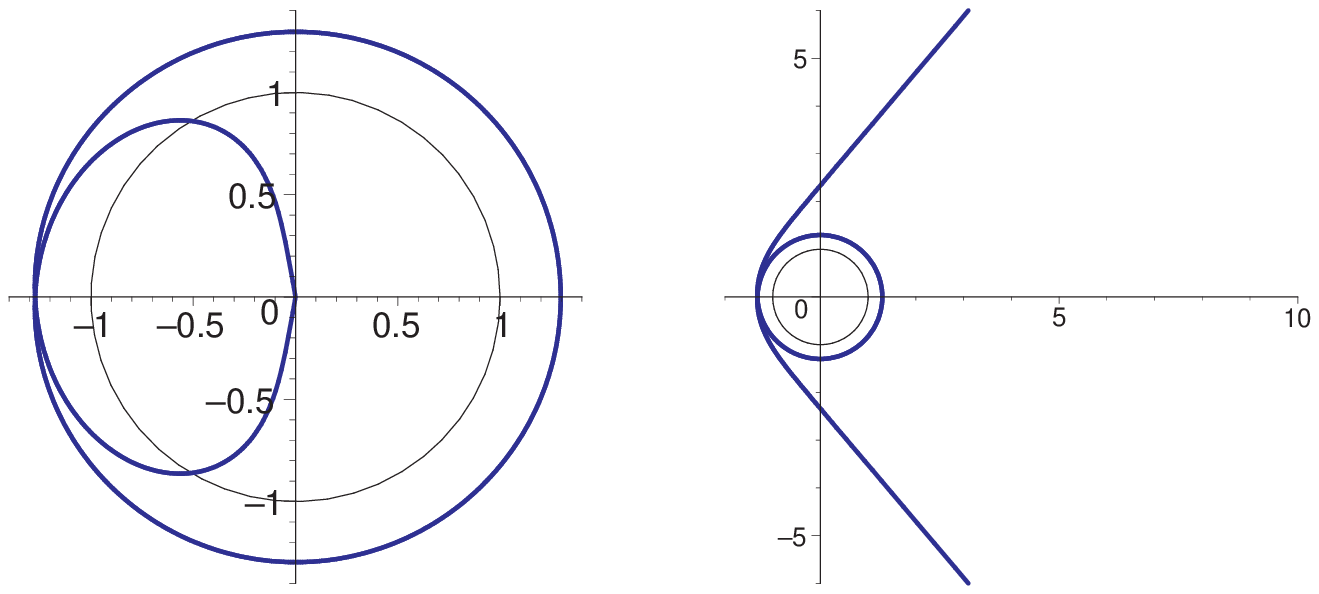}}
\subfigure[][$\mu=4.0$: terminating escape orbit]{\includegraphics[width=0.24\textwidth]{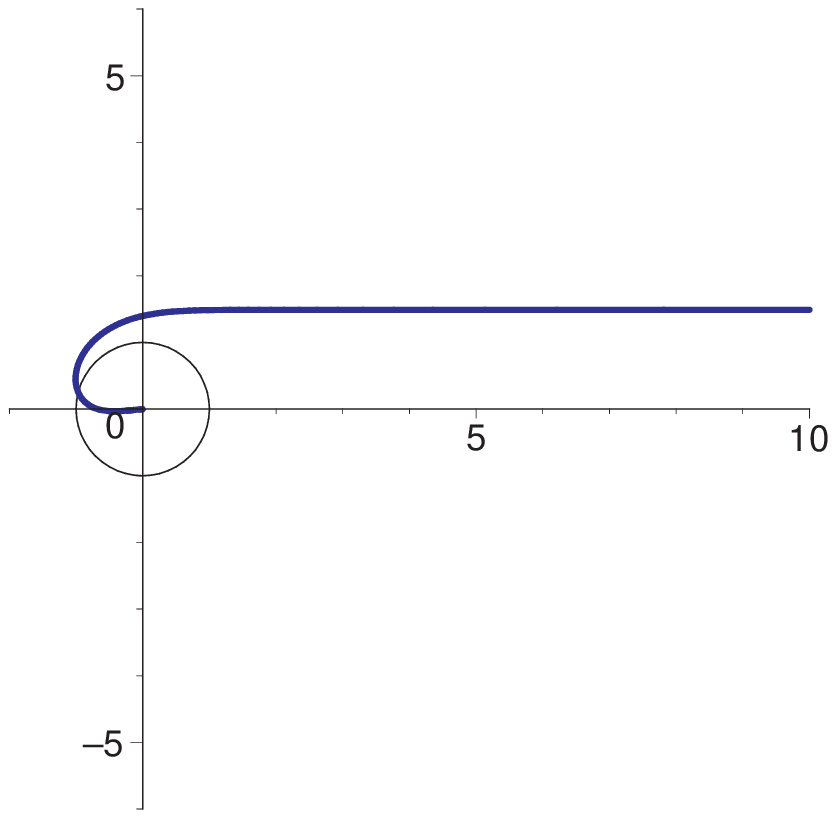}}
\end{center}
\caption{Orbits of a test particle in the 9--dimensional Schwarzschild space--time for $\lambda=0.15$. The black circle indicates the horizon. The plots are given in units of $r_s$.} \label{orbd9}
\end{figure}

The corresponding orbits of massive test particles are illustrated in Fig.~\ref{orbd9}. The chosen energies $\mu=2.6$ and $\mu=3.9$ are shown as dashed lines in the effective potential Fig.~\ref{sub:potS9d} which swap out the regions where a particle is allowed to move. For both chosen energy parameters terminating bound and escape orbits exist.  
For $\mu \geq 1$ also escape orbits become possible. For higher energies both kinds of orbits become more deformed and approach the unstable circular orbit corresponding to the maximum of the potential. Then terminating escape orbits arise.   

Let $e_{1}\dots e_{5}$ be the roots of the polynomial of 5th degree $P_{5}(u)$. 
The deflection angle for an escape orbit of a test particle in Schwarzschild space--times in 9 dimensions is
\begin{equation}
\label{defld9}
\Delta \varphi = 2 \int^{0}_{u_{e}} \frac{d u}{\sqrt{4 P_5(u)}} - \pi = \left\{ 
     \begin{array}{rl} 
            - 4 \omega_{11} - \pi & \text{if} \,\,\, \Re {e_{i}} < u_e \\
            - 4 \omega_{12} - \pi & \text{if} \,\,\, \Re {e_{i}} > u_e                               
     \end{array}      
\right.  \ ,
\end{equation}
where $u_e$ is a zero of $P_{5}(u)$ corresponding to an escape orbit and $\Re {e_{i}}$ is the real part of a corresponding complex root $e_i$.

\subsection{Schwarzschild space--time in 11 dimensions}\label{Sec:Schw_11D}

\begin{figure}[t]
\begin{center}
\subfigure[][$(\lambda,\mu)$--plot]{\label{sub:lamuS11d}%
\includegraphics[width=0.3\textwidth]{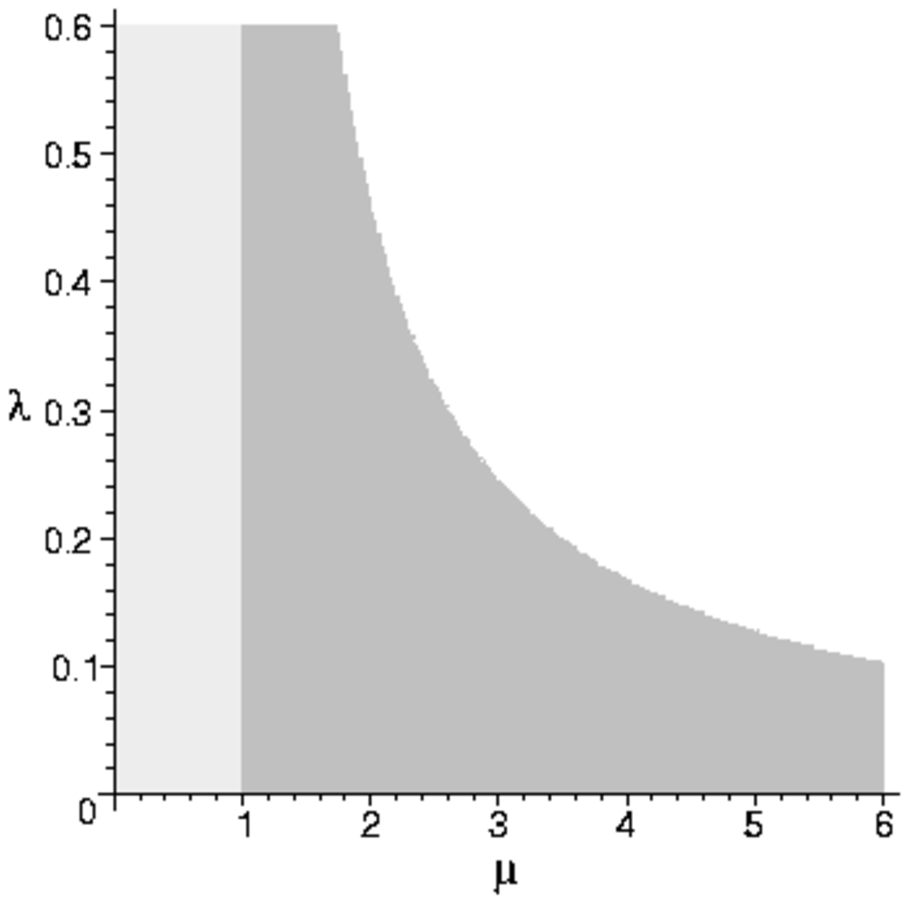}
} \quad %
\subfigure[][Potential for $\lambda=0.4$]{\label{sub:potS11d}%
\includegraphics[width=0.25\textwidth]{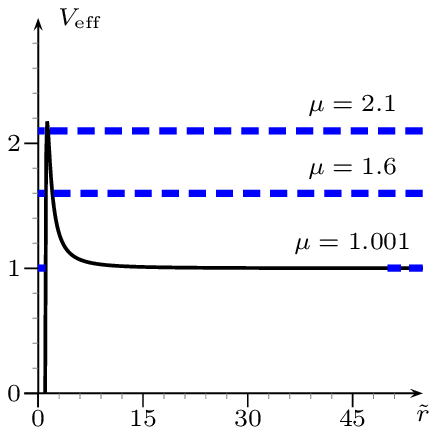}
}
\end{center}
\caption{11--dimensional Schwarzschild space-time. \subref{sub:lamuS11d} $(\lambda,\mu)$ plot (color code as in Fig.~{lamuSchw9D}). \subref{sub:potS11d} Effective potential. For the corresponding orbits see Fig.~\ref{orbSchwd11}. \label{lamuSchw11D}}
\end{figure}

For the Schwarzschild space--time in 11 dimensions the equation of motion~\eqref{EOMrphinormu} takes the form
\begin{equation}
\label{geodd11}
\left(\frac{d u}{d\varphi}\right)^2=4 u \left(u^5 + \lambda u^4 - u + \lambda(\mu-1)\right) = 4 P_6(u) \ .
\end{equation}
The number of zeros of this $P_6$ are shown in Fig.~\ref{lamuSchw11D} in the $(\lambda,\mu)$--plot together with the effective potential 
\begin{equation}
\label{Veff11S}
V_{\rm eff}= \left(1 - \frac{1}{\tilde{r}^8}\right) \left(1 + \frac{1}{\lambda\tilde{r}^2} \right) 
\end{equation}
which indicates no qualitative difference between the Schwarzschild space--times in 9 and 11 dimensions. With $z = \tilde{r}^2$ the extrema of the effective potential are given by $0 = z^4 - 4 \lambda z - 5$. The only possible real positive zero is given by the expression: $z=\frac{1}{\sqrt{2}}\left( \sqrt{k} + \left( -k + \frac{2\sqrt{2}\lambda}{\sqrt{k}} \right)^{\frac{1}{2}} \right)$ with $k=\frac{1}{3}\left(27\lambda^2+3\sqrt{81\lambda^4+375}\right)^{\frac{1}{3}}-5\left(27\lambda^2+3\sqrt{81\lambda^4+375}\right)^{-\frac{1}{3}}$ and describes a maximum.

\begin{figure}[t!]
\begin{center}
\subfigure[][$\mu=2.1$: terminating bound orbit and escape orbit]{\includegraphics[width=0.45\textwidth]{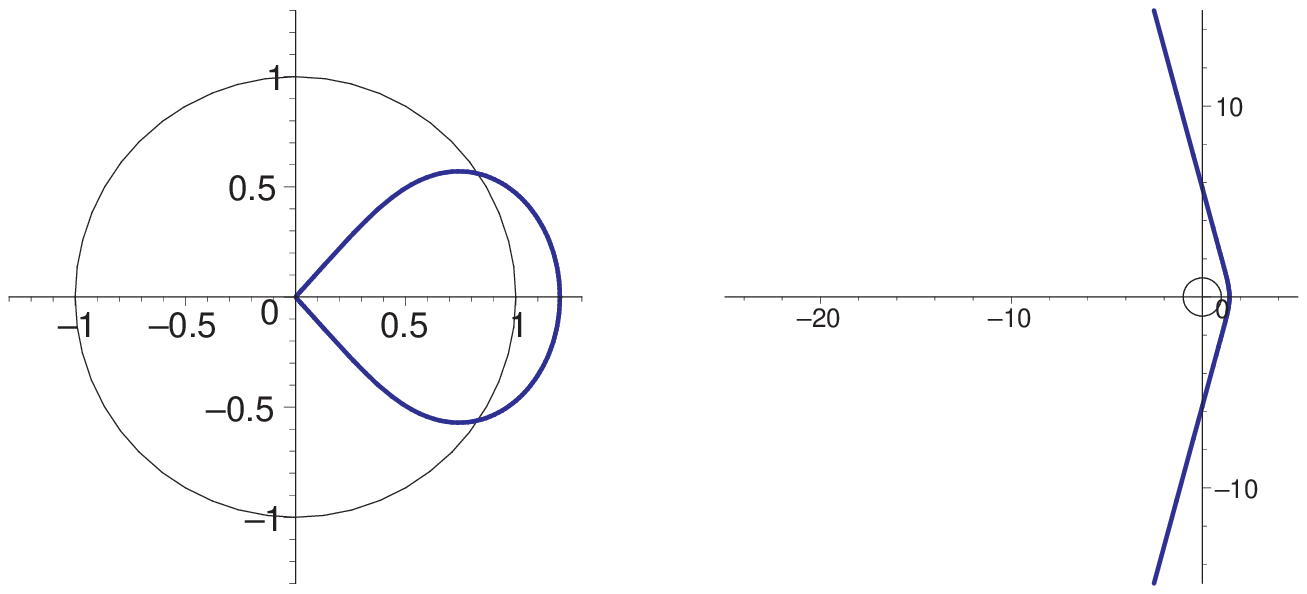}} 
\subfigure[][$\mu=2.1760084521$: terminating bound orbit and escape orbit]{\includegraphics[width=0.45\textwidth]{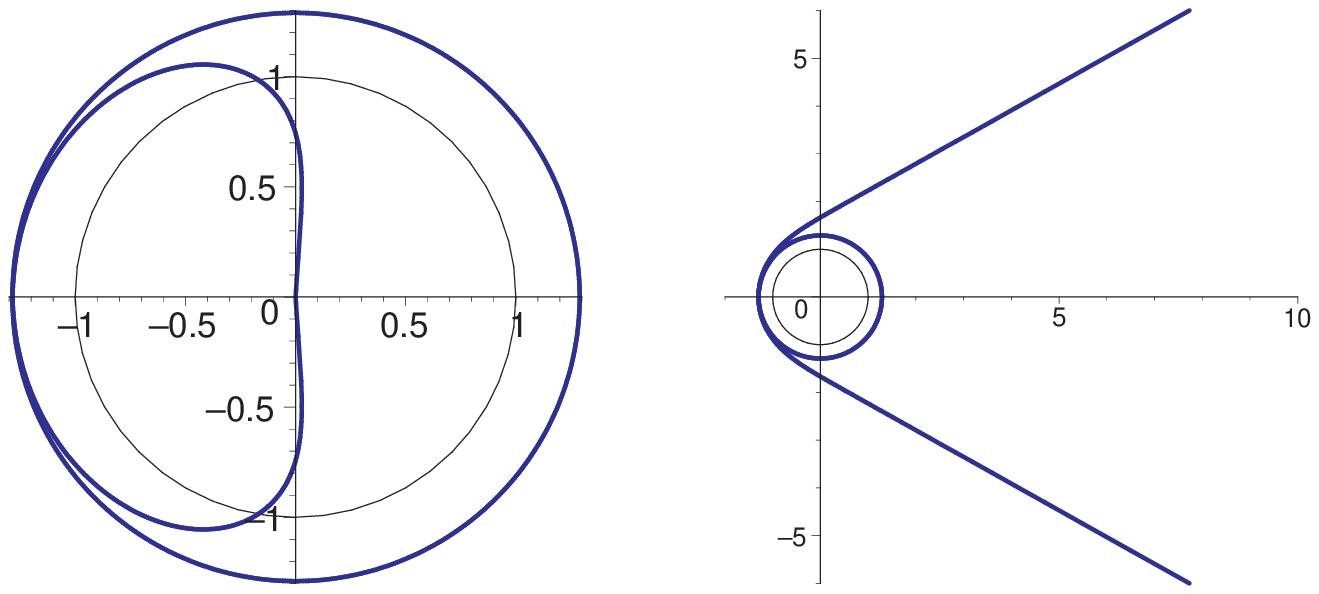}}
\end{center}
\caption{Orbits of a test particle in the Schwarzschild space--time in $11$ dimensions for $\lambda=0.4$. In (b) the energy is close to the energy of the unstable circular orbit related to the maximum of the effective potential in Fig.~\ref{lamuSchw11D}(b). \label{orbSchwd11}}
\end{figure}

There are no stable circular or periodic bound orbits.
A substitution $u=\frac{1}{x} + u_6$, where $u_6$ is a root of $P_6(u)$, transforms ~\eqref{geodd11} to $\left(x\frac{dx}{d\varphi}\right)^2= 4 P_5(x)$ which is of type~\eqref{rP_5a}. Then the solution of the geodesic equation for a test particle in an 11 dimensional Schwarzschild space--time is given by
\begin{equation}
\tilde{r}(\varphi) =  \frac{1}{\sqrt{- \frac{\sigma_2(\vec \varphi_{\infty, z_2})}{\sigma_1(\vec \varphi_{\infty, z_2})} + u_6}} \, ,  \label{solSchw11D}
\end{equation}
with the angular coordinate $\varphi_{z_2}$ from~\eqref{varphiP_5_physa} and $\vec\varphi_{\infty, z_2}$ from \eqref{varphi_infa}. $x_{\rm in}$ is again an initial angle which can be chosen to be a zero of $P_5(x)$. The orbits for various $\mu$ are illustrated in Fig.~\ref{orbSchwd11} and exhibit the same qualitative behavior as the orbits in 9 dimensions.

The deflection angle for an escape orbit of a test particle in the Schwarzschild space--time in 11 dimensions is
\begin{equation}
\label{defld11}
\Delta \varphi = 2 \int^{\infty}_{x_{e}} \frac{x dx}{\sqrt{4 P_5(x)}} - \pi = - 4 \omega_{21} - 4 \omega_{22} - \pi \ ,
\end{equation}
where $x_e$ is a zero of $P_{5}(x)$ corresponding to an escape orbit and we took $u_6=0$ here.

\section{Geodesics in higher dimensional Schwarzschild--(anti-)de Sitter space--times}\label{Sec:SchwdS}

We now consider timelike geodesics in Schwarzschild--(anti-)de Sitter (S(a)dS) space--time. A detailed discussion of geodesics in Schwarzschild--de Sitter space--time in 4 dimensions can be found in~\cite{HackmannLaemmerzahl08_PRD}. Here we restrict our considerations to 9 and 11 dimensions. The geodesic equations for 5 and 7 dimensions can be solved using the Weierstrass $\wp$--function.

\subsection{Schwarzschild--(anti-)de Sitter in 9 dimensions}

\begin{figure}[t!]
\begin{center}
\includegraphics[width=0.25\textwidth]{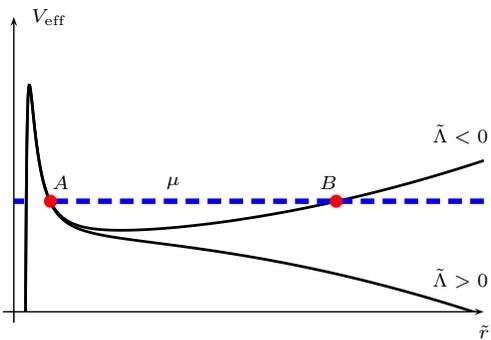}
\vspace{-0.5cm}
\caption{Potential for negative and positive cosmological parameter $\tilde{\Lambda}$. For the same value of a test particle energy $\mu$ it can be on a periodic bound orbit with perihelion and aphelion points $A$ and $B$ or on an escape one with a point of nearest approach $A$. \label{La+-}}
\end{center}
\end{figure}

For $d=9$ and $\eta=0$ we obtain from \eqref{EOMrphinormu} the equation of motion in the S(a)dS space--time in 9 dimensions
\begin{equation}
\label{dSgeod9d}
\left(\frac{du}{d\varphi}\right)^2= 4 \left(u^5 + \lambda u^4 - u^2 + \left(\lambda (\mu -1 ) + \frac{\tilde{\Lambda}}{28} \right) u + \frac{\tilde{\Lambda}}{28}  \lambda \right) = 4 P_5(u) \ ,
\end{equation}
which is of type~\eqref{rP_5b}. The solution of the geodesic equation in the $9$ dimensional Schwarzschild--(anti-)de Sitter space--time is of the form~\eqref{solSchw9D}.  
The sign of the cosmological constant is crucial for the form of the orbits. As it can be seen from the effective potential 
\begin{equation}
\label{Veff9SdS}
V_{\rm eff}= \left(1 - \frac{1}{\tilde{r}^6} - \frac{\tilde{\Lambda}}{28}\tilde{r}^2 \right) \left(1 + \frac{1}{\lambda\tilde{r}^2} \right) \,
\end{equation}
shown in Fig.~\ref{La+-}, for $\tilde{\Lambda} < 0$ a particle necessarily follows a bound orbit, while for $\tilde{\Lambda} > 0$ there are also escape orbits.

\subsubsection{Positive cosmological constant}\label{Sec:SchwdSposLambda}

\begin{figure}[t]
\begin{center}
\subfigure[][$(\lambda,\mu)$--plot, $\tilde{\Lambda}=8.7 \cdot 10^{-5}$]{\label{sub:lamuSdS9d+}%
\includegraphics[width=0.3\textwidth]{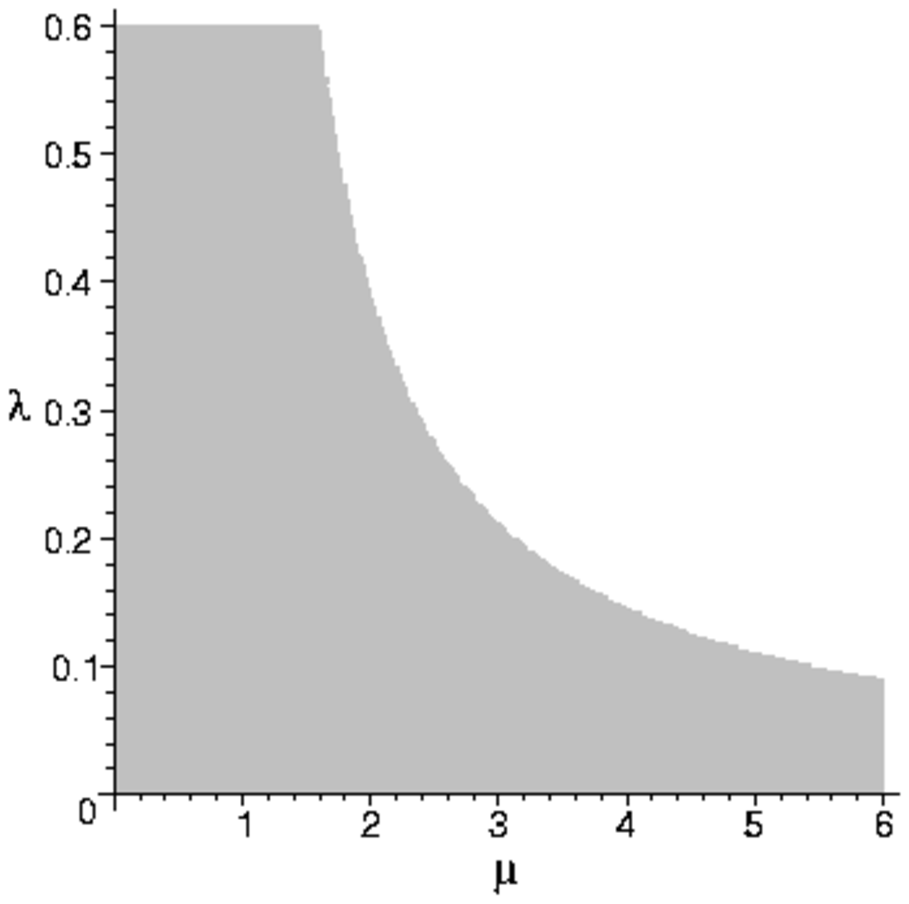}
} \quad %
\subfigure[][Potential for $\lambda=0.2$, $\tilde{\Lambda}=8.7 \cdot 10^{-5}$]{\label{sub:potSdS9d+}%
\includegraphics[width=0.25\textwidth]{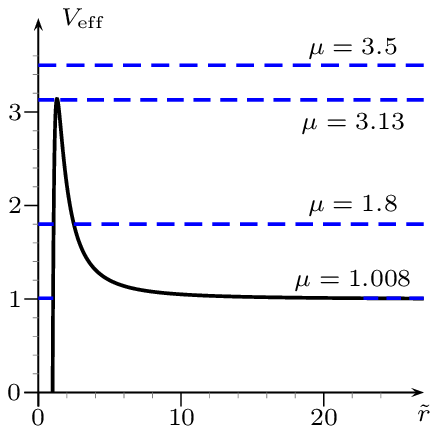}
}
\end{center}
\caption{9--dimensional SdS space--time: \subref{sub:lamuSdS9d+} $(\lambda,\mu)$--plot. \subref{sub:potSdS9d+} Effective potential. For the corresponding orbits see Fig.~\ref{dSorbd9+}. \label{lamuSchwdS9D+}}
\end{figure}

For $\tilde{\Lambda} > 0$ the positive zeros of $P_{5}(u)$ (and also of $P_{10}(\tilde{r})$) are shown in Fig.~\ref{sub:lamuSdS9d+}. From this Figure as well as from the effective potential $V_{\rm eff}$~\eqref{Veff9SdS} shown for $\lambda=0.2$ and $\tilde{\Lambda}=8.7\cdot 10^{-5}$ in Fig.~\ref{sub:potSdS9d+} 
it is clear that only terminating bound, terminating escape and escape orbits are possible. The qualitative structure is mainly the same as for the pure Schwarzschild case except that for small energies the escape orbits show reflection at the cosmological constant barrier, Fig.~\ref{sub:potSdS9d+}.

\begin{figure}[th!]
\begin{center}
\subfigure[][$\mu=1.008$: terminating bound orbit and escape orbit]{\includegraphics[width=0.45\textwidth]{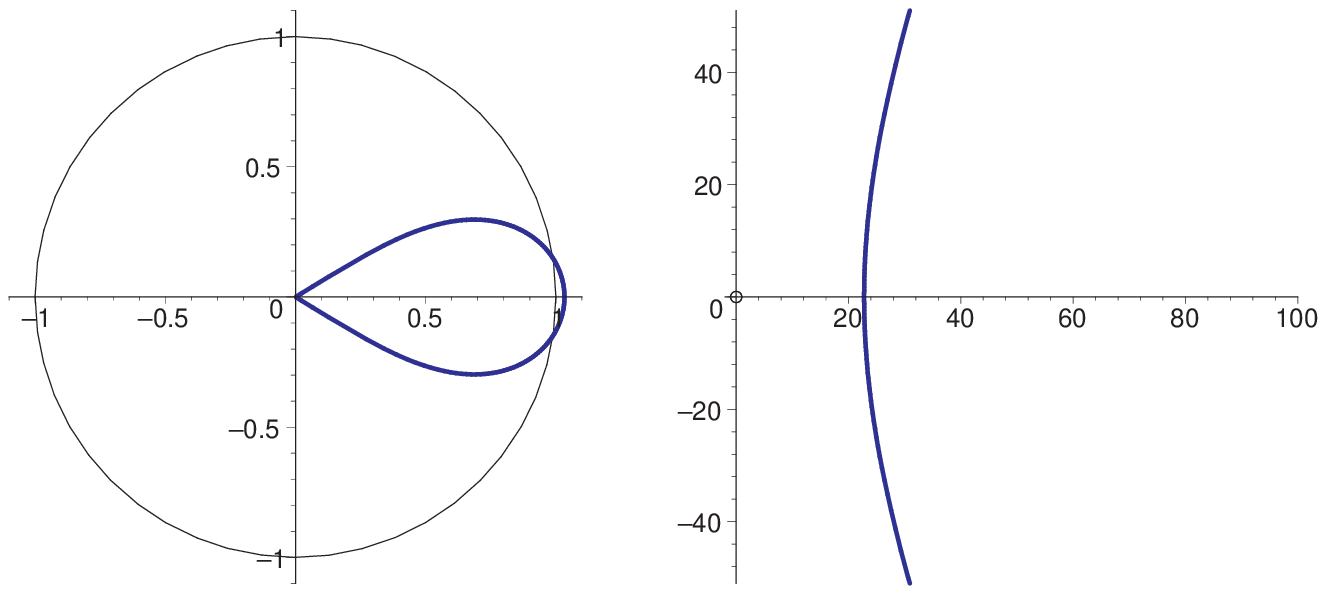}}
\subfigure[][$\mu=1.8$: terminating bound orbit and escape orbit]{\includegraphics[width=0.45\textwidth]{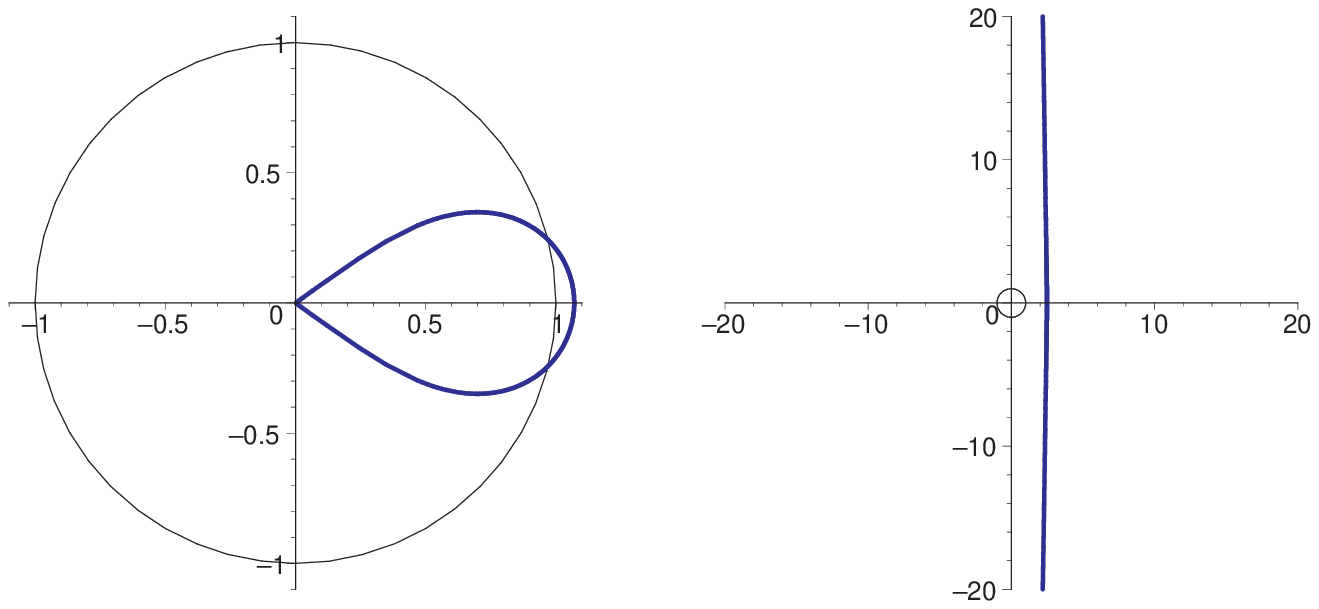}} 
\subfigure[][$\mu=3.13$: terminating bound orbit and escape orbit]{\includegraphics[width=0.45\textwidth]{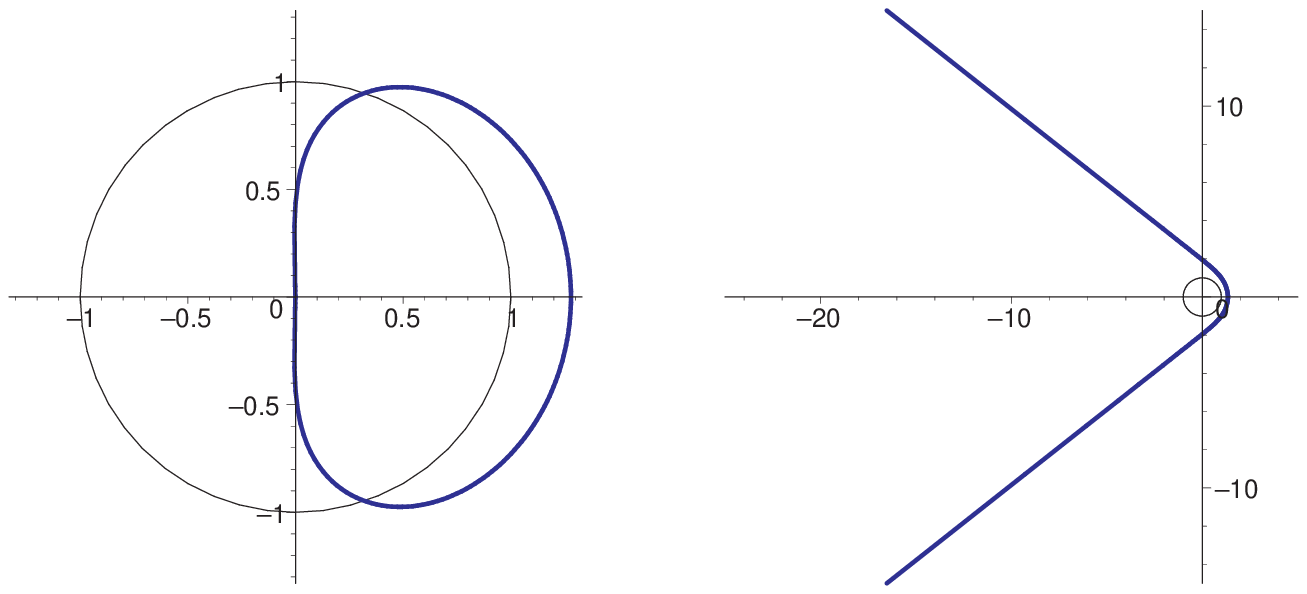}}
\subfigure[][$\mu=3.1392073469$: terminating bound orbit and escape orbit]{\includegraphics[width=0.45\textwidth]{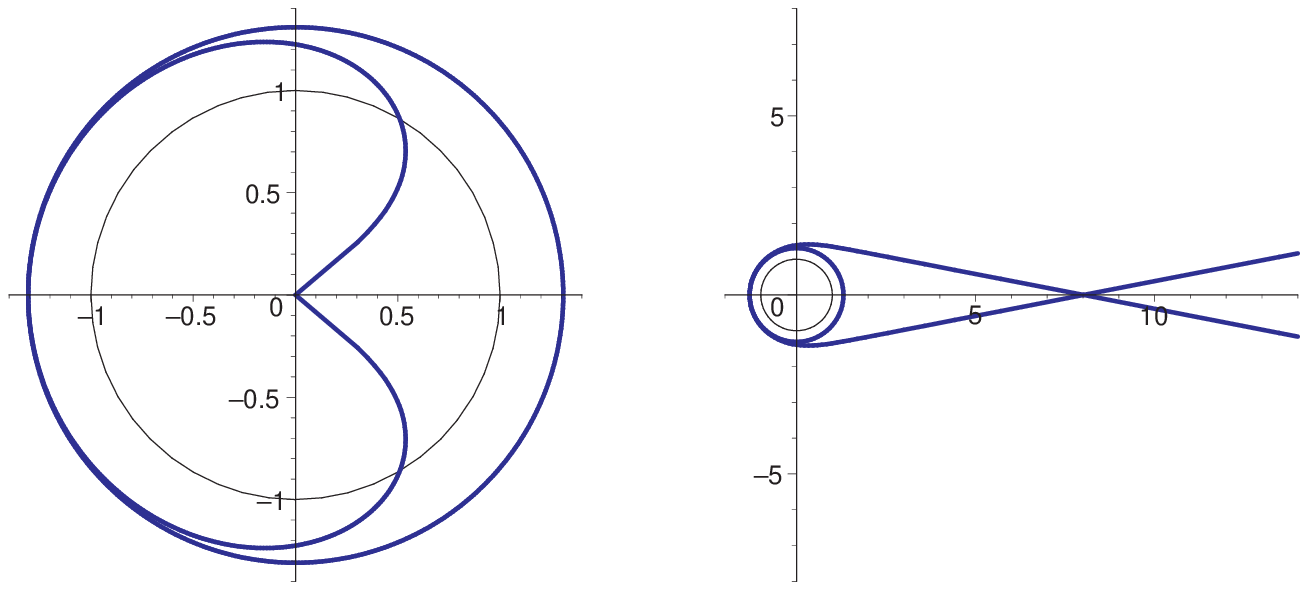}} 
\subfigure[][$\mu=3.13921$, terminating escape orbit]{\includegraphics[width=0.24\textwidth]{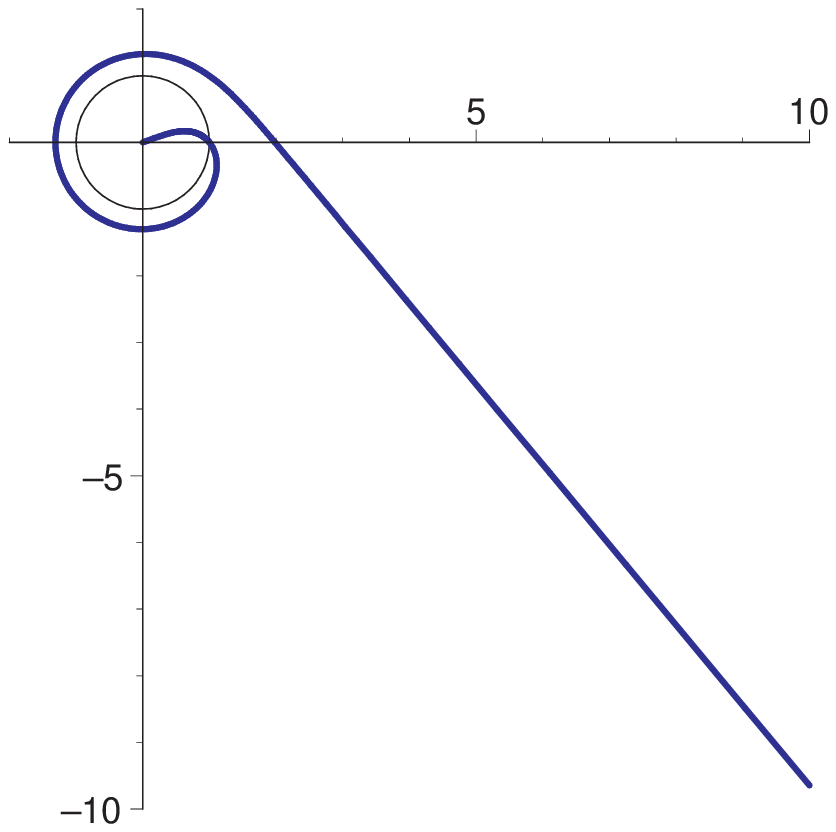}}
\subfigure[][$\mu=3.5$, terminating escape orbit]{\includegraphics[width=0.24\textwidth]{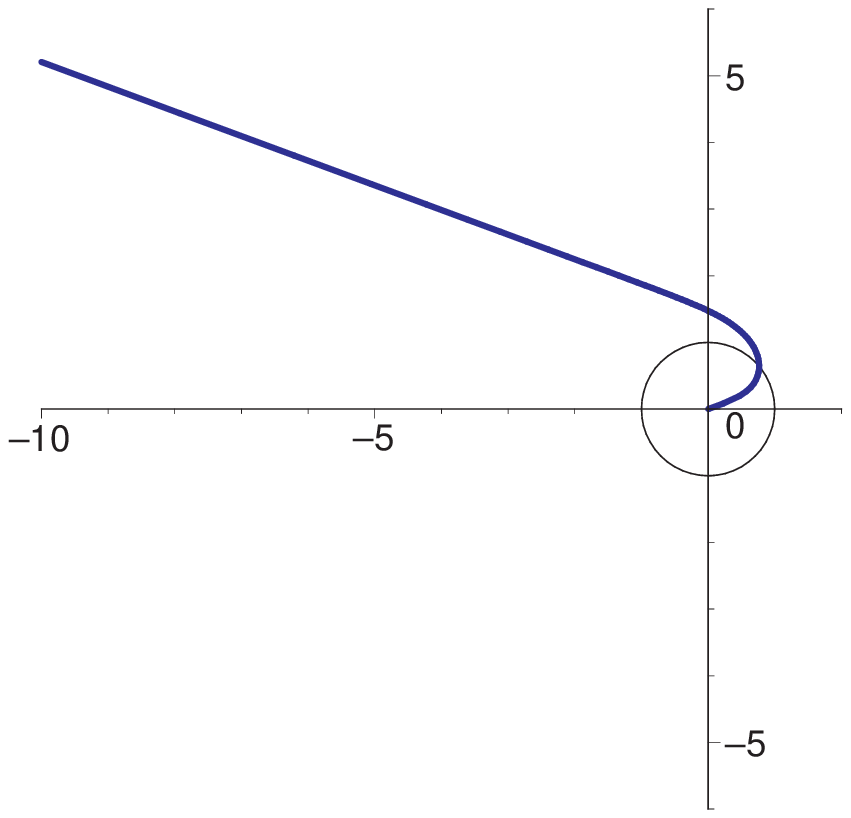}}
\end{center}
\caption{Examples of a test particle's motion in 9--dimensional SdS space--time with positive cosmological constant $\tilde{\Lambda}=8.7 \cdot 10^{-5}$ and $\lambda=0.2$. \label{dSorbd9+}}
\end{figure}

\subsubsection{Negative cosmological constant}

\begin{figure}[t]
\begin{center}
\subfigure[][$(\lambda,\mu)$--plot, $\tilde{\Lambda}= - 8.7 \cdot 10^{-5}$]{\label{sub:lamuSdS9d-}%
\includegraphics[width=0.3\textwidth]{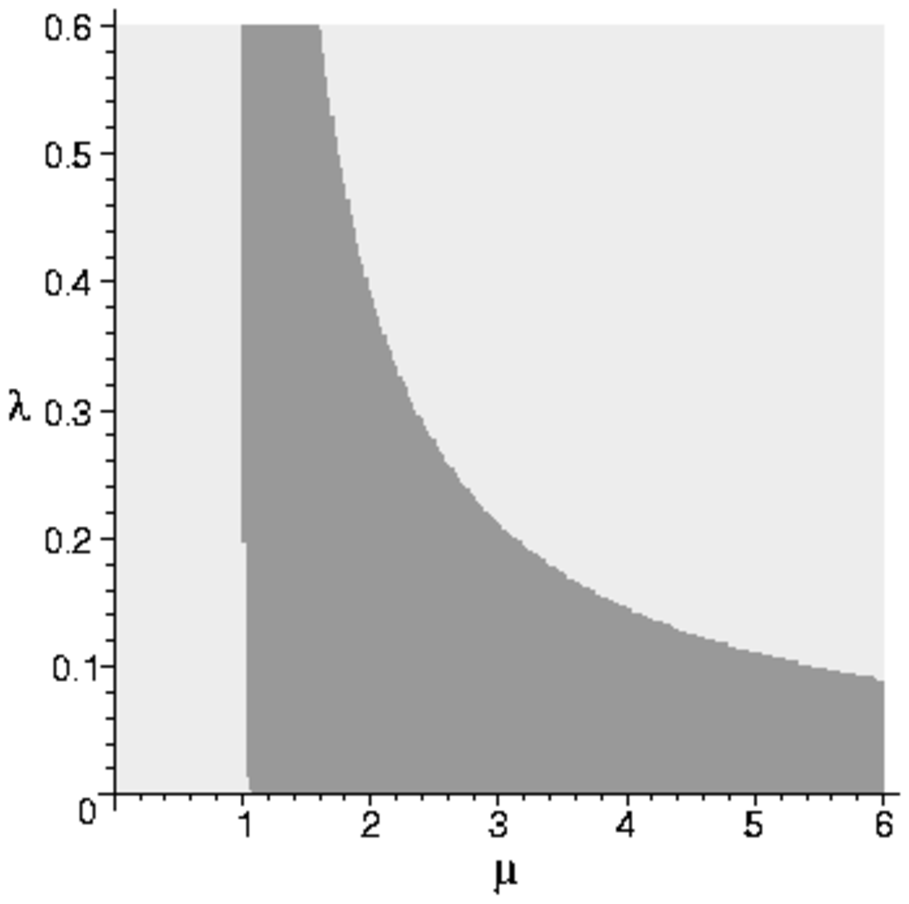}
} \quad %
\subfigure[][Potential for $\lambda=0.2$, $\tilde{\Lambda}= - 8.7 \cdot 10^{-5}$]{\label{sub:potSdS9d-}%
\includegraphics[width=0.25\textwidth]{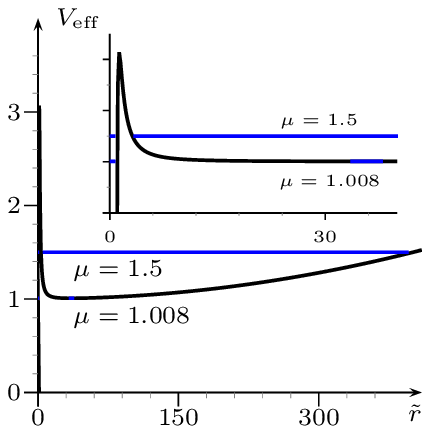}
}
\end{center}
\caption{9--dimensional SadS space--time: \subref{sub:lamuSdS9d-} $(\lambda,\mu)$--plot (here the dark gray color denotes 3 positive zeros). \subref{sub:potSdS9d-} Effective potential. For the corresponding orbits see Fig.~\ref{dSorbd9-}. \label{lamuSchwdS9D-}}
\end{figure}

\begin{figure}[t]
\begin{center}
\subfigure[][$\mu=1.008$: terminating bound and periodic bound orbits]{\includegraphics[width=0.45\textwidth]{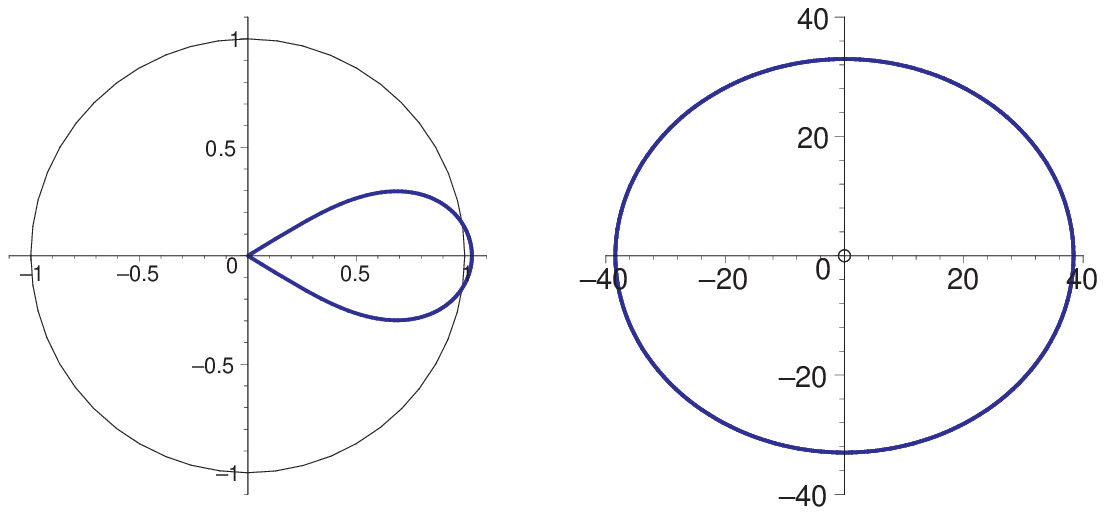}}
\subfigure[][$\mu=1.5$: terminating bound and periodic bound orbits]{\includegraphics[width=0.45\textwidth]{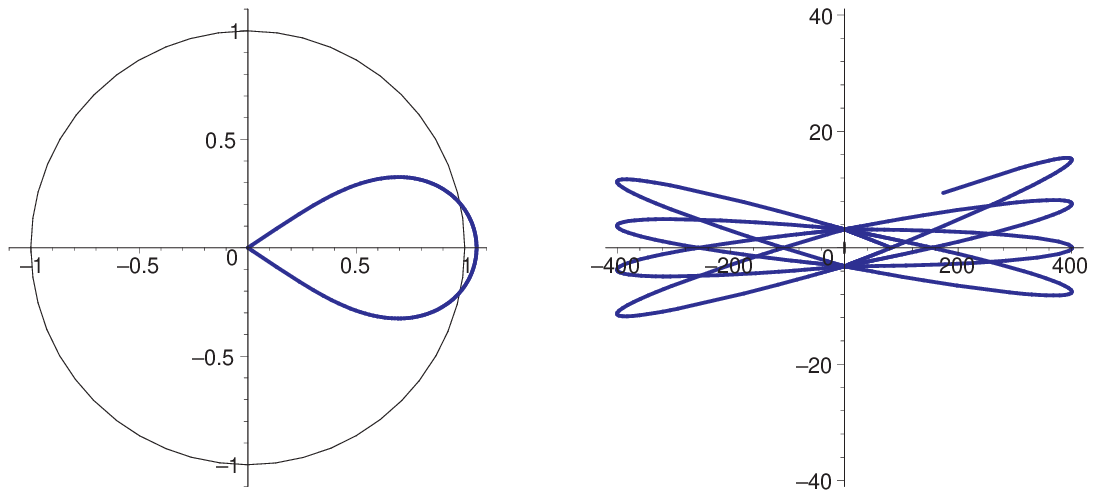}} 
\end{center}
\caption{Orbits of a test particle in 9--dimensional SadS space--time for $\tilde{\Lambda}= - 8.7 \cdot 10^{-5}$ and $\lambda=0.2$. \label{dSorbd9-}}
\end{figure}

For $\tilde{\Lambda} < 0$ the positive zeros of $P_{5}(u)$ are shown in  Fig.~\ref{sub:lamuSdS9d-}. For a negative cosmological constant the effective potential $V_{\rm eff}$ differs insofar as it tends to $+\infty$ for $r \rightarrow \infty$. Therefore each orbit is a bound orbit, some are periodic, some are terminating. 
This is related to the dark gray region in~Fig.~\ref{sub:lamuSdS9d-} corresponding to three real positive zeros giving one terminating bound orbit and one periodic bound orbit, and the light gray region indicating one real positive zero leading to terminating orbits (see Fig.~\ref{dSorbd9-}). 

For a periodic bound orbit restricted to the interval $[r_{\rm min}, r_{\rm max}]$ there is a perihelion shift 
\begin{equation}
\Delta_{\rm perihel}= 2 \pi - 2 \int_{e_i}^{e_{i+1}} \frac{du}{\sqrt{4P_5(u)}} = 2\pi - 4 \omega_{1 k} \ ,
\end{equation}
where $e_i$ and $e_{i+1}$ are that zeros of $P_5(u)$ related to $r_{\rm max}$ and $r_{\rm min}$ and the path $a_k$ surrounds the interval $[e_i, e_{i+1}]$. The actual $\omega_{1k}$ to be taken also depends on the position of the complex zeros of $P_5(u)$. 

\subsection{Schwarzschild--(anti-)de Sitter in 11 dimensions}

For $d = 11$ and $q = 0$ we obtain from \eqref{EOMrphinormu} the equation of motion 
\begin{equation}
\label{dSgeod11d}
\left(\frac{du}{d\varphi}\right)^2= 4 \Bigg(u^6 + \lambda u^5 - u^2 + \left(\lambda (\mu -1 ) + \frac{\tilde{\Lambda}}{45} \right) u + \frac{\tilde{\Lambda}}{45} \lambda \Bigg) = 4 P_6(u) \,
\end{equation}
and from~\eqref{Veff} the effective potential in the S(a)dS space--time in 11 dimensions
\begin{equation}
\label{Veff11SdS}
V_{\rm eff}= \left(1 - \frac{1}{\tilde{r}^8} - \frac{\tilde{\Lambda}}{45}\tilde{r}^2 \right) \left(1 + \frac{1}{\lambda\tilde{r}^2} \right) \ .
\end{equation}
In analogy to the case of 9 dimensions we infer from~\eqref{dSgeod11d} that $P_6(u)$ has two or no real positive roots, thus, leading either to escape or to bound terminating orbits.

\begin{figure}[t]
\begin{center}
\subfigure[][$(\lambda,\mu)$--plot, $\tilde{\Lambda}=8.7 \cdot 10^{-5}$]{\label{sub:lamuSdS11d}%
\includegraphics[width=0.3\textwidth]{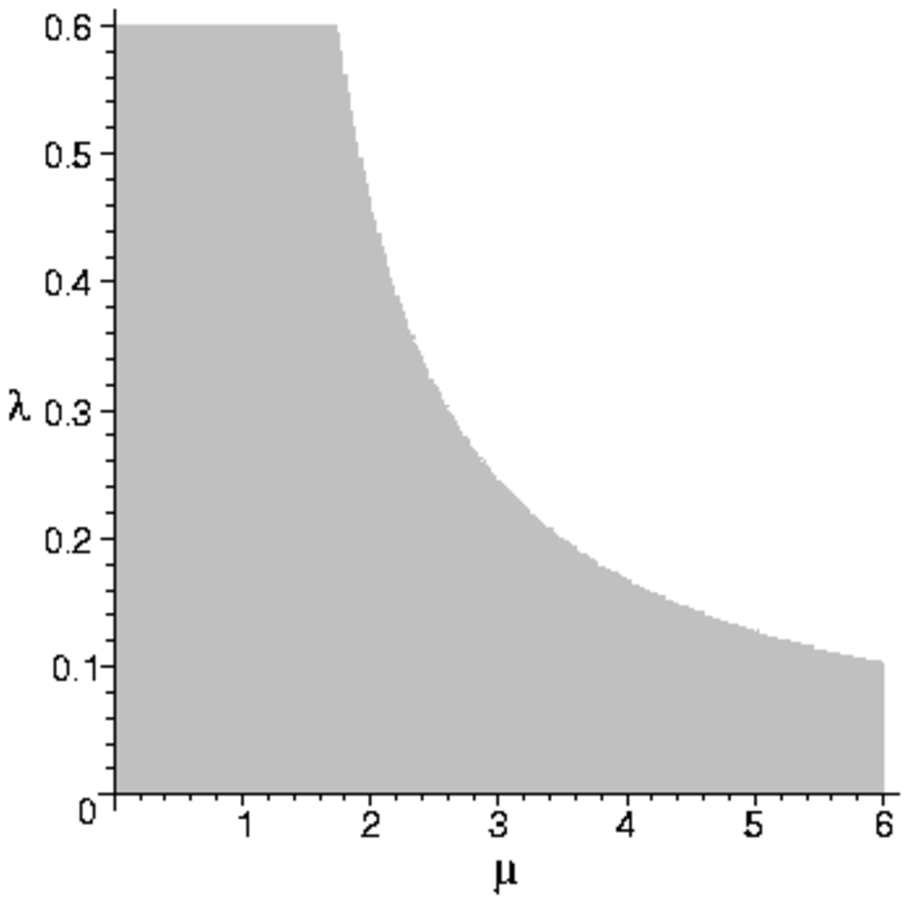}
} \quad %
\subfigure[][Potential for $\lambda=0.15$, $\tilde{\Lambda}=8.7 \cdot 10^{-5}$]{\label{sub:potSdS11d}%
\includegraphics[width=0.25\textwidth]{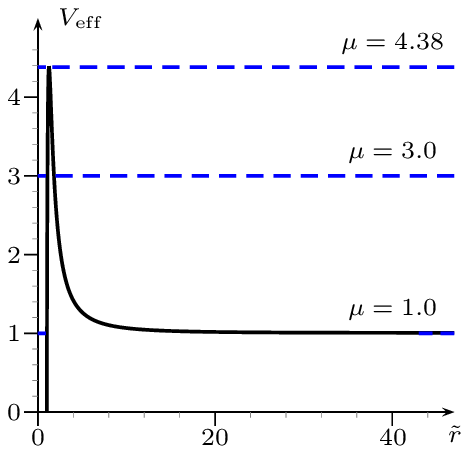}
}
\end{center}
\caption{11 dimensional S(a)dS space--time: \subref{sub:lamuSdS11d} $(\lambda,\mu)$ plot (color code as above). \subref{sub:potSdS11d} Effective potential. For the corresponding orbits see Fig.~\ref{dSorbd11+}.  \label{lamuSchwdS11D+}}
\end{figure}

\begin{figure}[th!]
\begin{center}
\subfigure[][$\mu=1.0$: terminating bound orbit and escape orbit]{\includegraphics[width=0.45\textwidth]{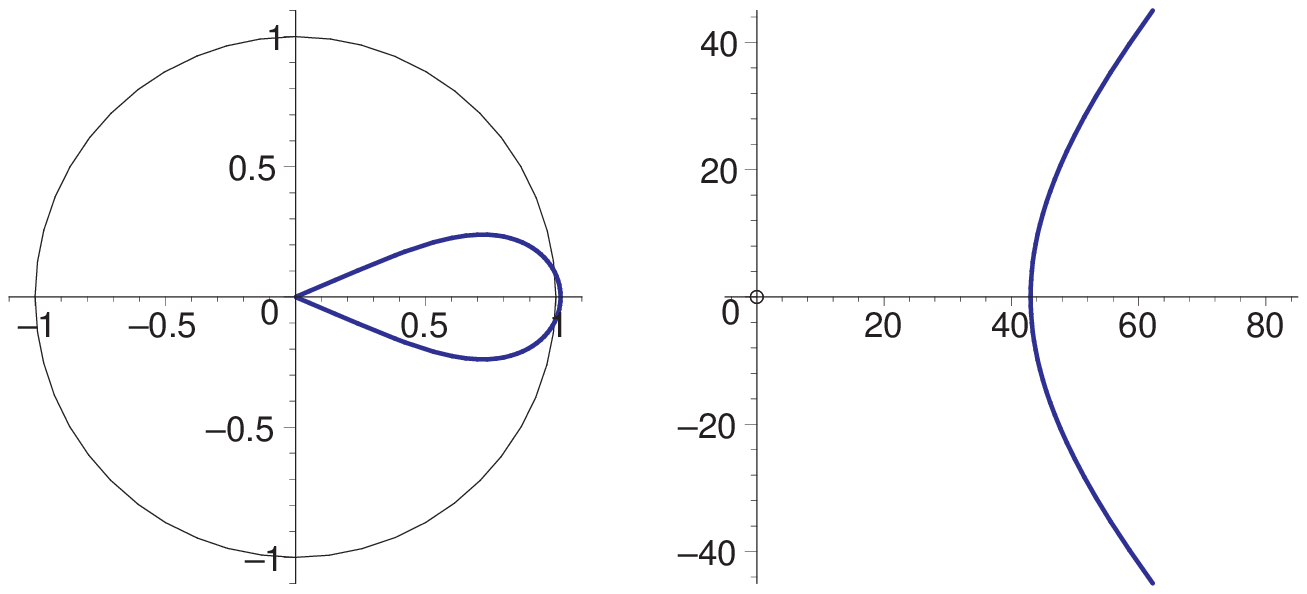}}
\subfigure[][$\mu=4.38$: terminating bound orbit and escape orbit]{\includegraphics[width=0.45\textwidth]{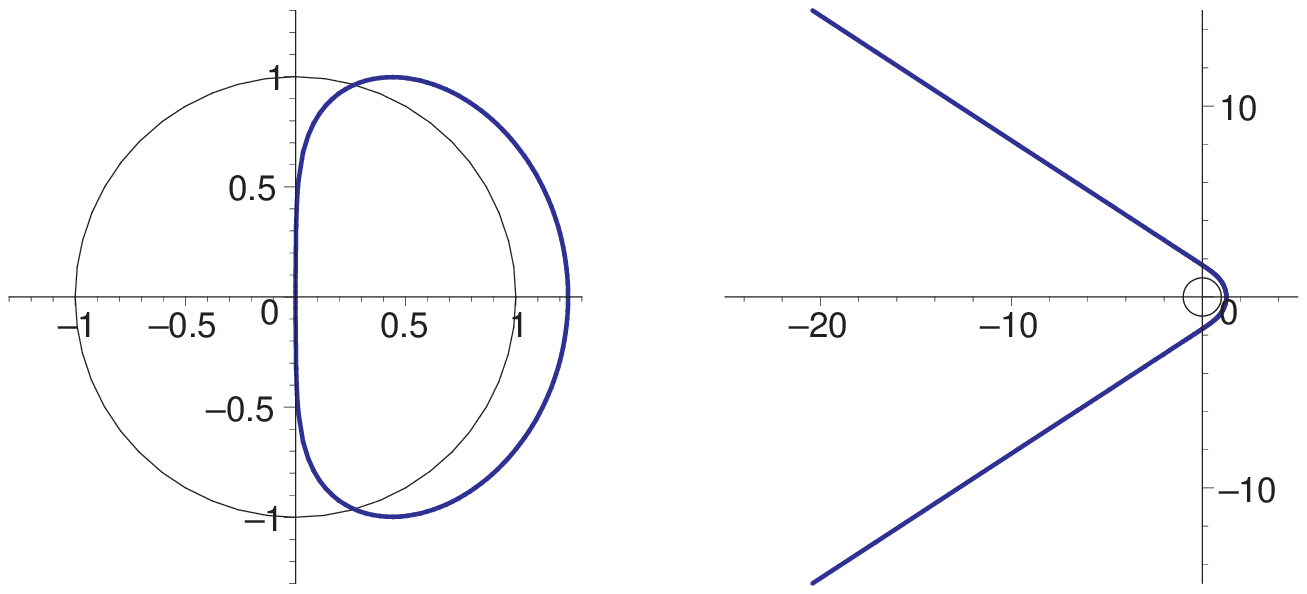}} \\
\end{center}
\caption{Geodesics of a test particle in in 11--dimensional S(a)dS space--time with positive cosmological constant $\tilde{\Lambda}=8.7 \cdot 10^{-5}$ and $\lambda=0.15$. \label{dSorbd11+}}
\end{figure}

For obtaining the analytical solution of \eqref{dSgeod11d} we have to introduce a new variable $x$ by $u=\frac{1}{x} + u_6$, where $u_6$ is a zero of $P_6(u)$. This transforms  equation~\eqref{dSgeod11d} containing $P_6(u)$ to a new equation $\left(x\frac{dx}{d\varphi}\right)^2= 4 P_5(x)$ which is of type~\eqref{rP_5a} and possesses a solution of the form \eqref{solSchw11D}. The obtained orbits show no qualitative differences to the 9 dimensional SdS case, see Figs.~\ref{lamuSchwdS11D+} and \ref{dSorbd11+} for the plots for positive cosmological constant.

\section{Geodesics in Reissner--Nordstr\"om space-times in 7 dimensions}\label{Sec:RN}

\begin{figure}[t]
\begin{center}
\subfigure[][$(\lambda,\mu)$--plot, $\eta=0.7$]{\label{lamuRN7d}%
\includegraphics[width=0.3\textwidth]{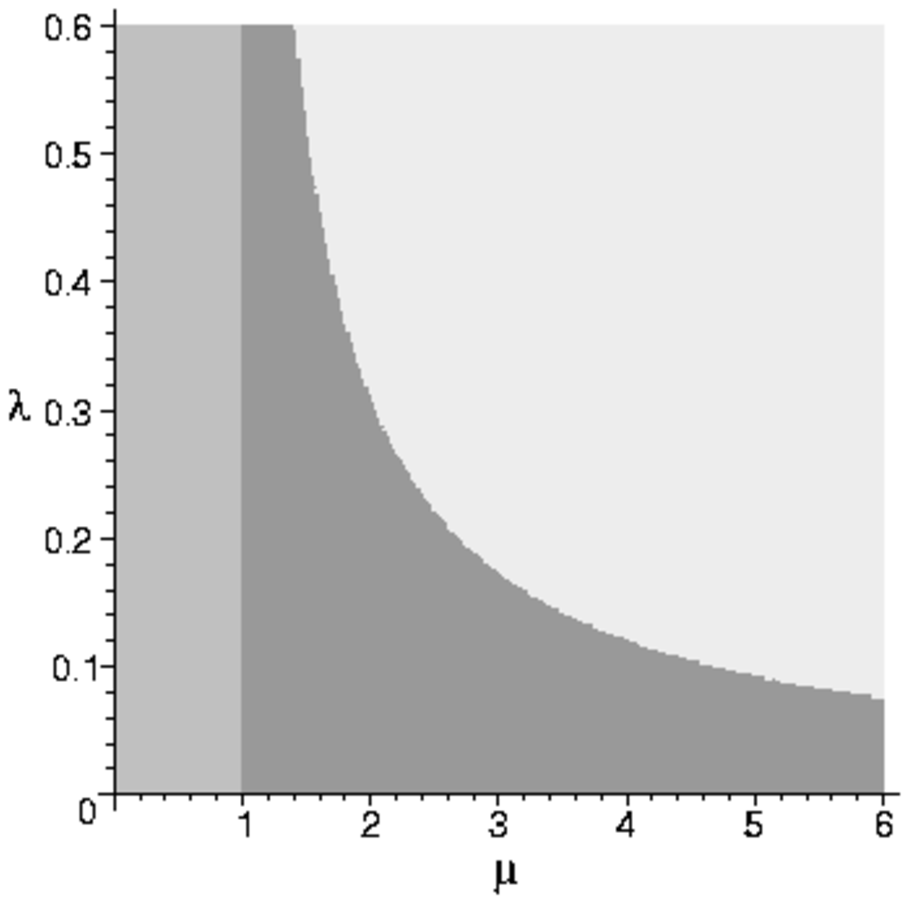}
} \quad %
\subfigure[][Potential for $\lambda=0.35$, $\eta=0.7$]{\label{potRN7d}%
\includegraphics[width=0.25\textwidth]{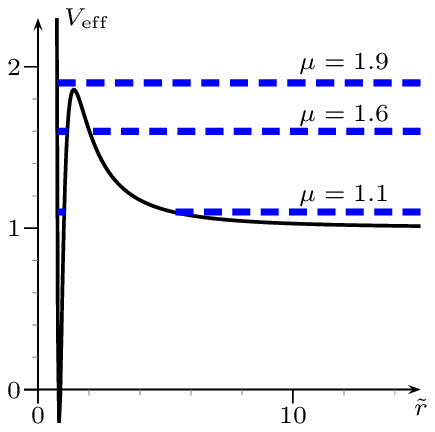}
}
\end{center}
\caption{7--dimensional RN space--time: \subref{lamuRN7d} $(\lambda,\mu)$--plot (gray scales as above). \subref{potRN7d} Effective potential for $\eta=0.7$. For the corresponding orbits see Fig.~\ref{RNorbd7}. \label{lamuRN7Deta07}}
\end{figure}

Let us now address the higher dimensional Reissner--Nordstr\"om (RN) space--time and investigate the effects of the charge parameter $\eta$ on the timelike geodesics. We restrict ourselves to $d=7$ of~\eqref{EOMrphinormu}
\begin{equation}
\label{RNgeod7d}
\left(\frac{du}{d\varphi}\right)^2= 4 u \left( - \eta^4 u^5 - \eta^4 \lambda u^4 + u^3 +  \lambda u^2 - u + \lambda (\mu -1 )   \right) = 4 P_6(u) \ .
\end{equation}
The relation of the zeros of $P_6(u)$ with the parameters $\lambda$ and $\nu$ are shown in   
Fig.~\ref{lamuRN7d} for $\eta=0.7$.
From this plot as well as from the effective potential 
\begin{equation}
\label{Veff7RN}
V_{\rm eff}= \left(1 - \frac{1}{\tilde{r}^4} +  \frac{\eta^4}{\tilde{r}^8} \right) \left(1 + \frac{1}{\lambda\tilde{r}^2} \right) \,
\end{equation}
shown in 
Fig.~\ref{potRN7d} it is clear that we have three types of orbits: For one zero (light gray area) we have an escape orbit, for two zeros (gray area) we have periodic bound orbit, and for three zeros (dark gray area) we have a periodic bound orbit and escape orbit. The relative maximum in $V_{\rm eff}$ shown in Fig.~\ref{potRN7d} increases for larger $L$ (smaller $\lambda$) and the relative minimum decreases for smaller $q$ (larger $\eta$). 

In order to obtain explicitly the analytic solution we have to substitute $u = - \frac{1}{x} + u_6$, where $u_6$ is the root of $P_6(u)$, what transforms \eqref{RNgeod7d} to $\left(x\frac{dx}{d\varphi}\right)^2 = 4 P_5(x)$. Then the solution of the geodesic equation in Reissner--Nordstr\"om space--time in $7$ dimensions is given by the equation 
\begin{equation}
\tilde{r}(\varphi) =  \frac{1}{\sqrt{ \frac{\sigma_2(\vec \varphi_{\infty, z_2})}{\sigma_1(\vec \varphi_{\infty, z_2})} + u_6}} \, ,  \label{solRN7D}
\end{equation}
with the angular coordinate $\varphi_{z_2}$ from~\eqref{varphiP_5_physa} and $\vec\varphi_{\infty, z_2}$ from \eqref{varphi_infa}. $x_{\rm in}$ again is an initial position which can be chosen to be a zero of $P_5(x)$.

RN space--times have two distinctive features. The first is the existence of two horizons, an inner (Cauchy) and an outer horizon which for $d=7$ are given by 
\begin{equation}
\label{hor7RN}
r_{\rm inner}= \frac{1}{\sqrt{2}}\left(2-2\sqrt{1-4\eta^4}\right)^\frac{1}{4} \, \qquad r_{\rm outer}= \frac{1}{\sqrt{2}}\left(2+2\sqrt{1-4\eta^4}\right)^\frac{1}{4}  \, ,
\end{equation}
and the second is an antigravitating potential barrier preventing a particle from falling into the singularity which is related to $\lim_{\tilde r \rightarrow 0} V_{\rm eff} = + \infty$. For $\eta=\frac{1}{\sqrt{2}}$ the two horizons coincide and we obtain a degenerate RN space--time. In this case the extrema of $V_{\rm eff}$ in \eqref{Veff7RN} are given by $\tilde{r}_1 = 2^{- 1/4}$ and $\tilde{r}_2 = \sqrt{\lambda+\sqrt{\lambda^2+ \frac{5}{2}}}$. At $\tilde{r}_1$ the effective potential possesses a minimum with $V_{\rm eff}(\tilde{r}_1)=0$, and at $\tilde{r}_2$ a maximum. That means in particular that a particle with energy $\mu=0$ moves on a circular orbit with radius $\tilde{r} = r_{\rm inner} = r_{\rm outer}$ corresponding to the two coinciding horizons. For $0 < \mu < 1$ a particle necessarily follows a periodic bound orbit. For $1 < \mu < V_{\rm eff}(\tilde{r}_2)$ we have a periodic bound and an escape orbit and for $\mu > V_{\rm eff}(\tilde{r}_2)$ only escape orbits.

\begin{figure}[th!]
\begin{center}
\subfigure[][$\mu=1.01$, $\eta=0.2$: many--world periodic bound orbit and escape orbit]{\label{orbRN7d:a}%
\includegraphics[width=0.45\textwidth]{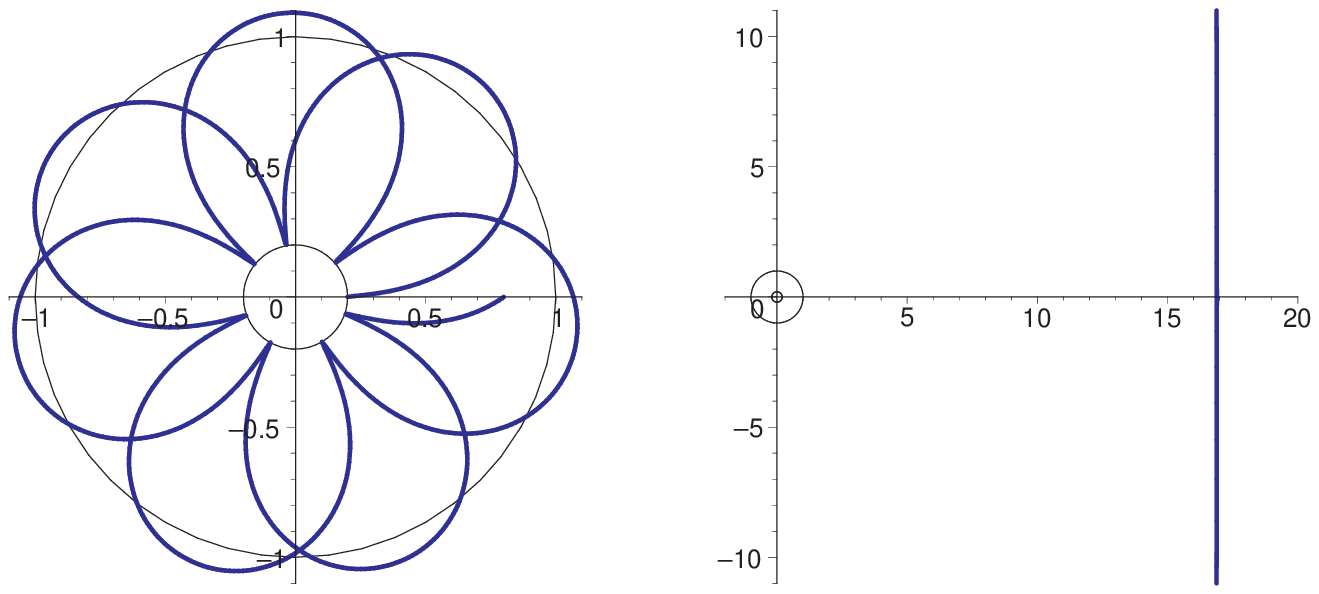}} \quad
\subfigure[][$\mu=1.82$, $\eta=0.2$: many--world periodic bound orbit and escape orbit]{\label{orbRN7d:b}%
\includegraphics[width=0.45\textwidth]{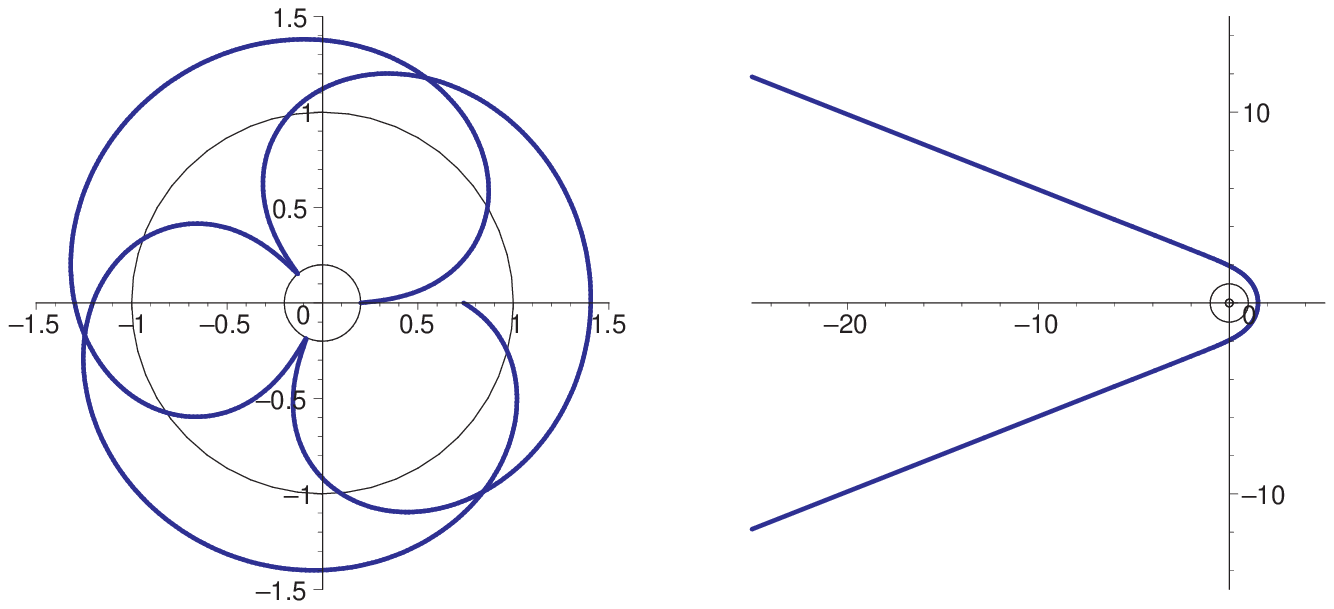}} \\
\subfigure[][$\mu=1.1$, $\eta=0.7$: many--world periodic bound orbit and escape orbit]{\label{orbRN7d:c}%
\includegraphics[width=0.45\textwidth]{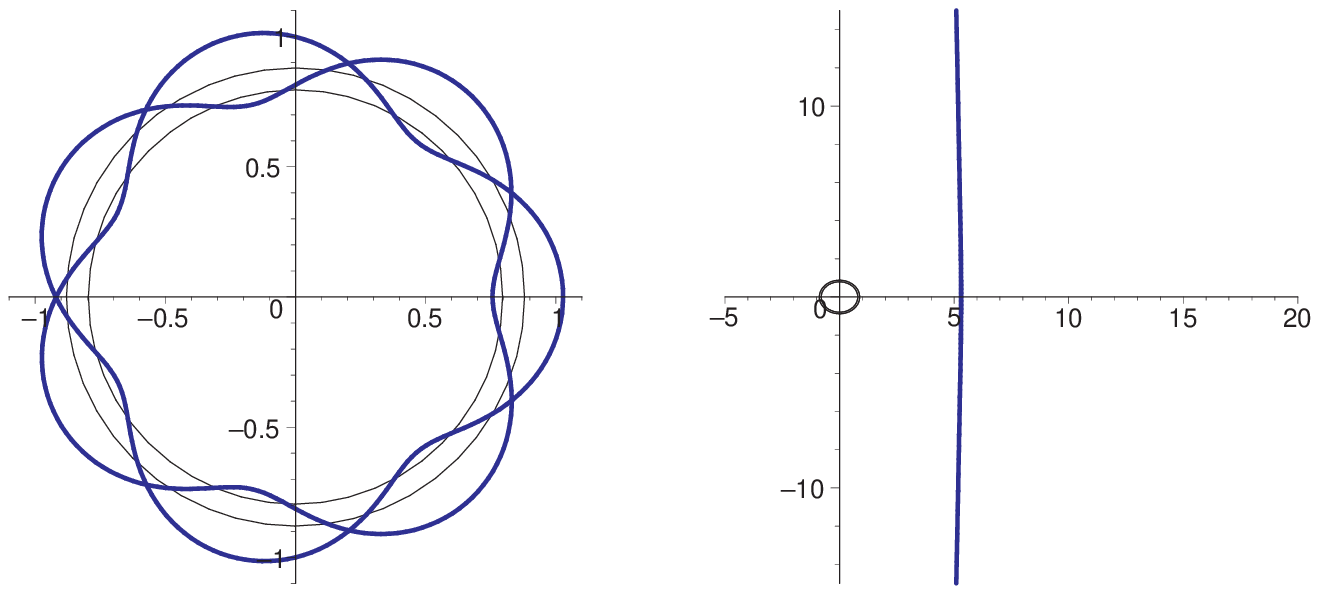}} \quad
\subfigure[][$\mu=1.6$, $\eta=0.7$: many--world periodic bound orbit and escape orbit]{\label{orbRN7d:d}%
\includegraphics[width=0.45\textwidth]{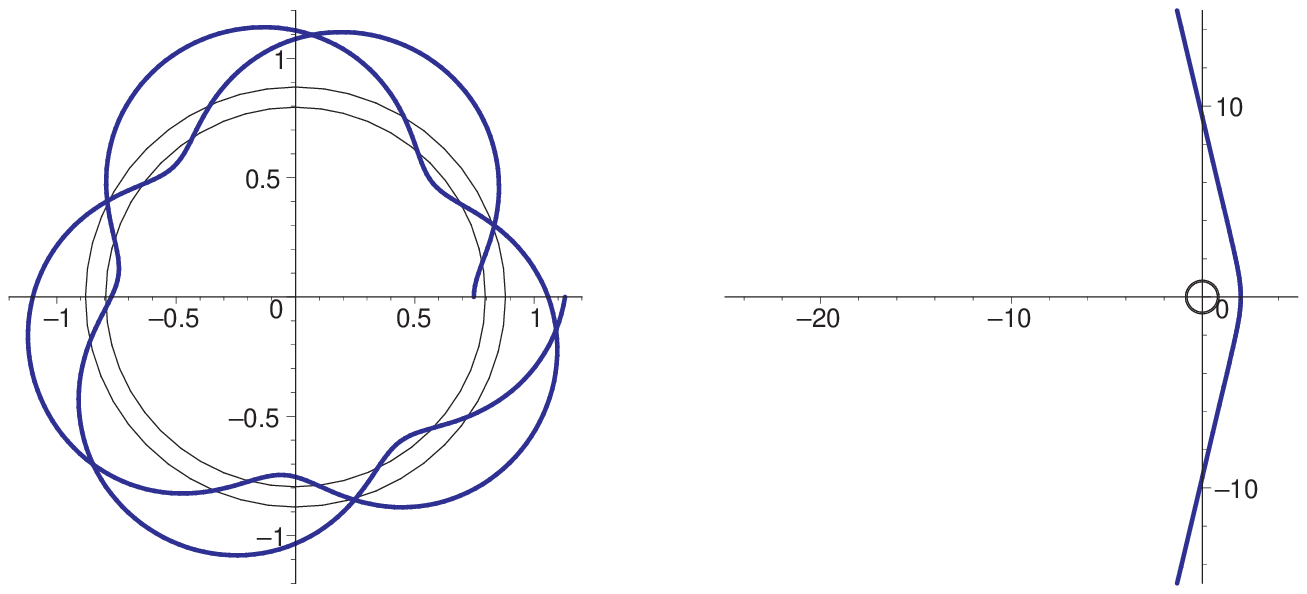}}\\
\subfigure[][$\mu=1.9$, $\eta=0.2$, two--world escape orbit]{\label{orbRN7d:e}%
\includegraphics[width=0.19\textwidth]{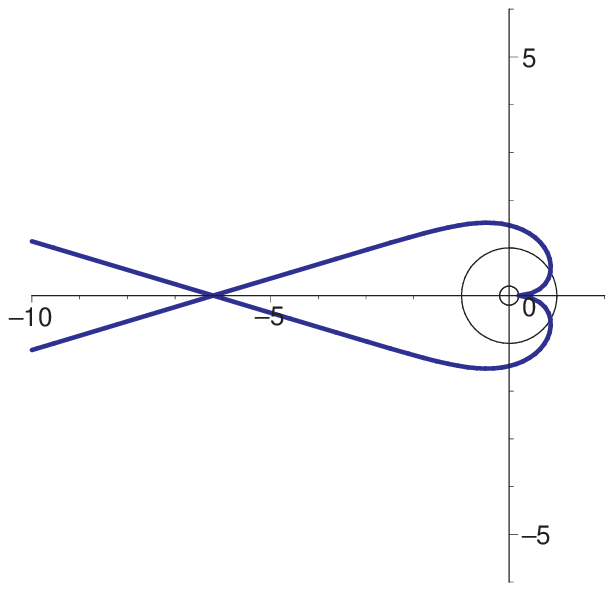}} %
\subfigure[][$\mu=1.9$, $\eta=0.7$: two--world escape orbit]{\label{orbRN7d:f}%
\includegraphics[width=0.19\textwidth]{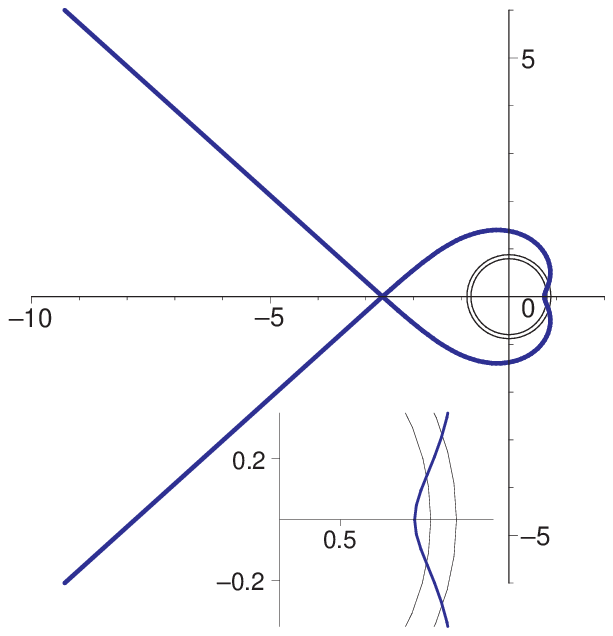}}   
\end{center}
\caption{Orbits for a test particle motion in Reissner--Nordstr\"om space--time in $7$ dimensions for $\lambda=0.35$. \label{RNorbd7}}
\end{figure}

For arbitrary $\eta$ the extrema of $V_{\rm eff}$ are defined by $0 = z^4 - 2 z^3 \lambda - 3 z^2 + 4 \eta^4 \lambda z + 5 \eta^4$ with $z = \tilde{r}^2$. Using again Descartes, there are two or no extrema at positive $z$. For two extrema there is a range of $\mu$ for which there are periodic bound as well as escape orbits, and a range of $\mu$ with escape orbits only, see the plots in Fig.~\ref{RNorbd7}. If there is no extremum there are escape orbits only.

A particle moving along one of the periodic bound orbits shown in Fig.~\ref{RNorbd7} crosses, say, from the outside, first the outer and then the inner horizon, both indicated by black circles. After crossing the inner horizon the particle is reflected by the charge induced potential barrier and again crosses the Cauchy horizon in the opposite direction, thereby entering a new copy of the RN space--time. This can be inferred from the Carter--Penrose diagram of the RN space-time shown, e.g., in \cite{MTW73} or \cite{Chandrasekhar83}. By proceeding further along its $\tilde{r}$--periodic motion, the particle again and again enters new copies of the RN space--time within its analytic continuation. This may be called a {\it many--world periodic bound orbit}. Since $E^2 \geq 0$, all bound orbits have to cross both horizons. 

The same happens with the escape orbits shown in Fig.~\ref{orbRN7d:e},\subref{orbRN7d:f}. The incoming particle crosses both horizons (see the inset in Fig.~\ref{orbRN7d:f}), is reflected by the potential barrier and, by crossing again the horizons in the transverse direction, enters a new copy of the RN space--time. Analogously this may be called a {\it two--world escape orbit}. Such a scattering or flyby makes the spacecraft disappear into another universe.

The perihelion shift for periodic bound orbits is
\begin{equation}
\Delta_{\rm perihel}= 2 \pi - 2 \int_{e_i}^{e_{i+1}} \frac{xdx}{\sqrt{4P_5(x)}} = 2 \pi - 4 \omega_{2 k} \ ,
\end{equation}
where the zeros $e_i$ and $e_{i+1}$ of $P_5(x)$ are again related to the range of motion $[r_{\rm min}, r_{\rm max}]$ and the path $a_k$ surrounds the interval $[e_i, e_{i+1}]$. Again, the actual $\omega_{2k}$ depends on the position of the other zeros of $P_5$ under consideration. Due to the fact that the motion does not take place within one universe it is not clear how one could measure this perihelion shift. One possibility might be that an astronaut adds up the angles $\varphi$ measured in the various universes starting from the position of largest distance with respect to the gravitating body to the next position of largest distance. In this case only the astronaut can carry through the measurement and only he or she has access to the result. This kind of effect may be called a ``many--word perihelion shift''.

\section{Geodesics in Reissner--Nordstr\"om--(anti-)de Sitter space-times}\label{Sec:RNdS}

As a last example of the motion in higher dimensional space--times, we discuss RN-de Sitter space--times in $4$ and $7$ dimensions. This is the most compound case due to a large number of parameters which includes the mass and the charge of the gravitating source and the cosmological constant. Owing to its complexity, we start the investigation for 4 dimensions and then consider 7 dimensions.

\subsection{Reissner--Nordstr\"om--(anti-)de Sitter in 4 dimensions}\label{Sec:RNdS4D}

\begin{figure}[t!]
\begin{center}
\subfigure[][$(\lambda,\mu)$--plot, $\tilde{\Lambda}=8.7 \cdot 10^{-5}$, $\eta=0.1$]{\label{sub:lamuRNdS4d}%
\includegraphics[width=0.3\textwidth]{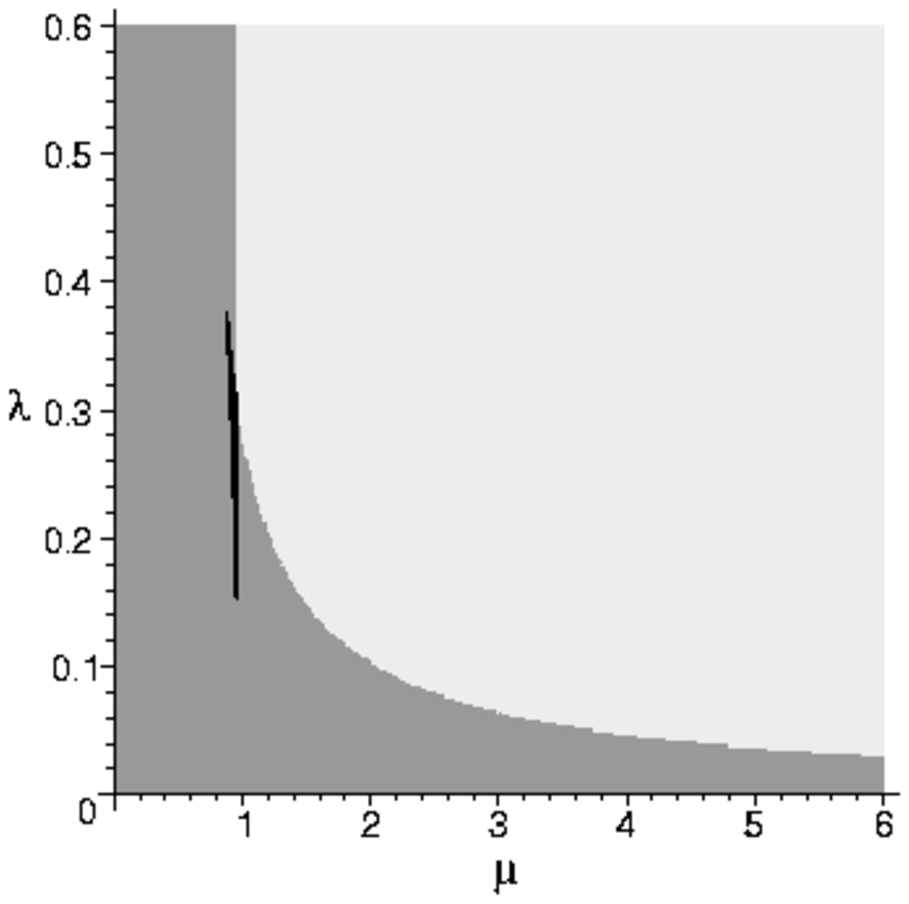}
}
\quad
\subfigure[][Potential for $\lambda=0.1$]{\label{sub:potRNdS4d}%
\includegraphics[width=0.25\textwidth]{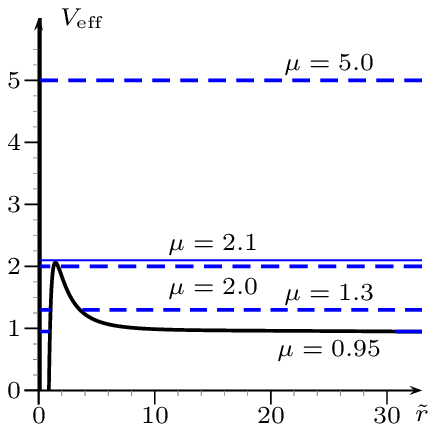}
}
\quad\quad %
\subfigure[][Potential for $\lambda=0.3576404$]{\label{sub:potRNdS4d_2}%
\includegraphics[width=0.25\textwidth]{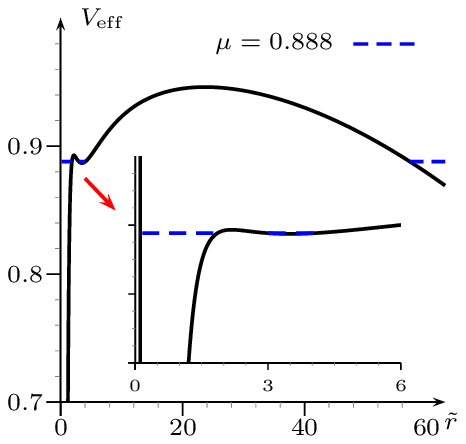}
}  
\end{center}
\caption{7--dimensional RN--(anti-)de Sitter space--time: \subref{sub:lamuRNdS4d} $(\lambda,\mu)$--plot (the black region indicates $5$ positive roots, dark gray--$3$ and light gray--$1$). \subref{sub:potRNdS4d} Effective potential for $\eta=0.1$ and $\tilde{\Lambda}=8.7 \cdot 10^{-5}$. For the corresponding orbits see Figs.~\ref{RNdSorbd4} and \ref{RNdStwoBOUND}.  \label{lamuRNdS4D}}
\end{figure}

Since already the solution of the geodesic equation in 4--dimensional S(a)dS space--times requires the application of hyperelliptic integrals \cite{HackmannLaemmerzahl08_PRL,HackmannLaemmerzahl08_PRD} the same is true for the geodesic equation in 4--dimensional RN--(anti-)de Sitter space-times 
\begin{equation}
\label{RNdSgeod4d}
\left(\frac{d\tilde{r}}{d\varphi}\right)^2= \frac{1}{3} \Bigg( \tilde{\Lambda} \lambda \tilde{r}^6 + \left(3 \lambda (\mu - 1) + \tilde{\Lambda} \right) \tilde{r}^4 + 3 \lambda \tilde{r}^3  -3 (\eta \lambda +1 ) \tilde{r}^2 + 3 \tilde{r} - 3 \eta \Bigg) = \frac{1}{3} P_6(\tilde{r}) \, .
\end{equation}
The zeros of~\eqref{RNdSgeod4d} are represented in a $(\lambda,\mu)$--plot Fig.~\ref{sub:lamuRNdS4d}. This is in line with the form of the effective potential 
\begin{equation}
\label{Veff4RNdS}
V_{\rm eff}= \left(1 - \frac{1}{\tilde{r}} - \frac{\tilde{\Lambda}}{3}\tilde{r}^2 +  \frac{\eta}{\tilde{r}^2} \right) \left(1 + \frac{1}{\lambda\tilde{r}^2} \right) \,
\end{equation}
shown in Fig.~\ref{lamuRNdS4D}(b) for $\lambda=0.1$, $\eta=0.1$, $\tilde{\Lambda}=8.7\cdot 10^{-5}$. 

\begin{figure}[th!]
\begin{center}
\subfigure[][$\mu=0.95$: many--world periodic bound orbit and escape orbit]{\label{orbRNdS4d:a}%
\includegraphics[width=0.45\textwidth]{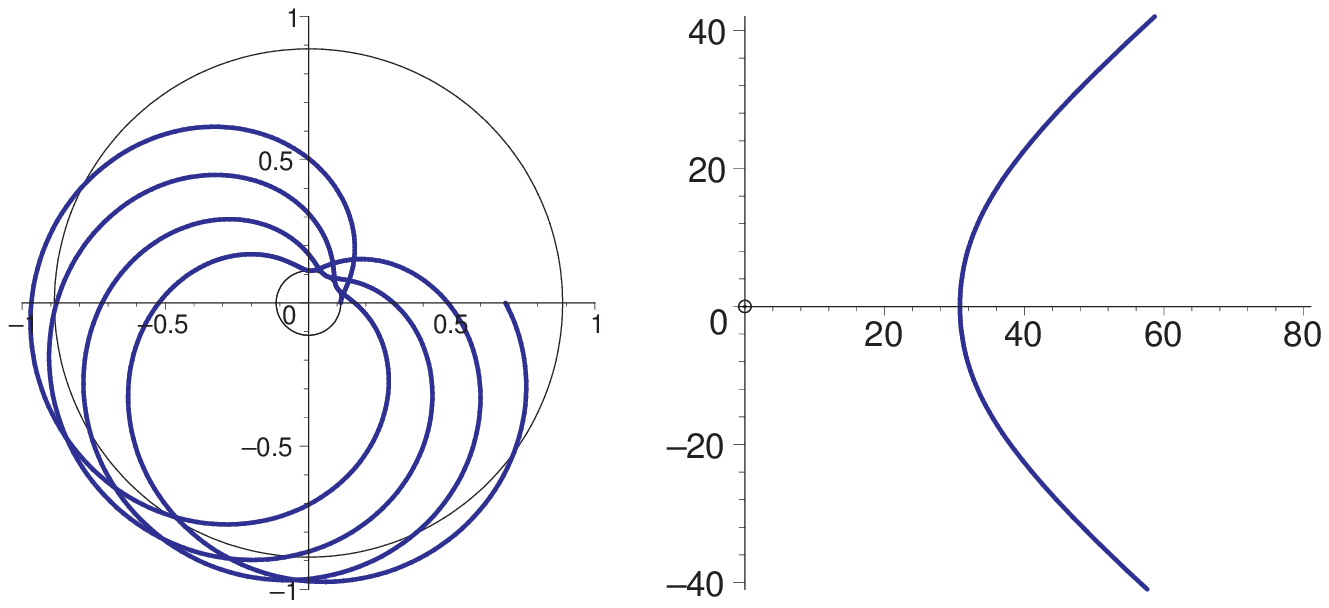}}
\subfigure[][$\mu=1.3$: many--world periodic bound orbit and escape orbit]{\label{orbRNdS4d:b}%
\includegraphics[width=0.45\textwidth]{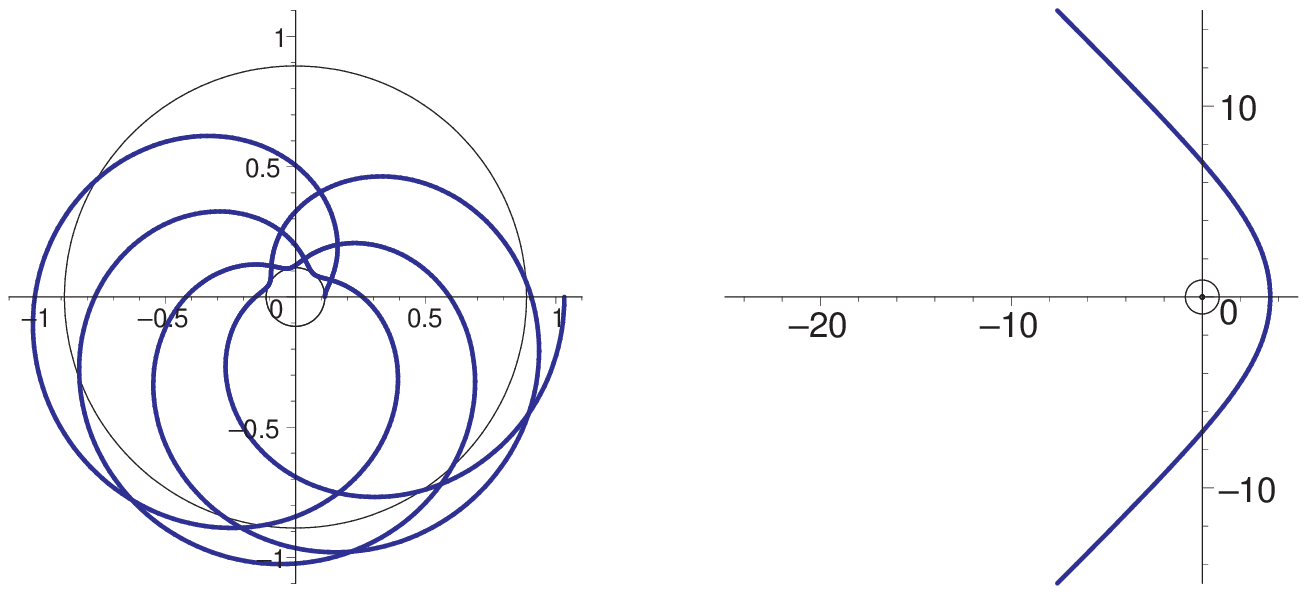}} \\ 
\subfigure[][$\mu=2$: many--world periodic bound orbit and escape orbit]{\label{orbRNdS4d:c}%
\includegraphics[width=0.45\textwidth]{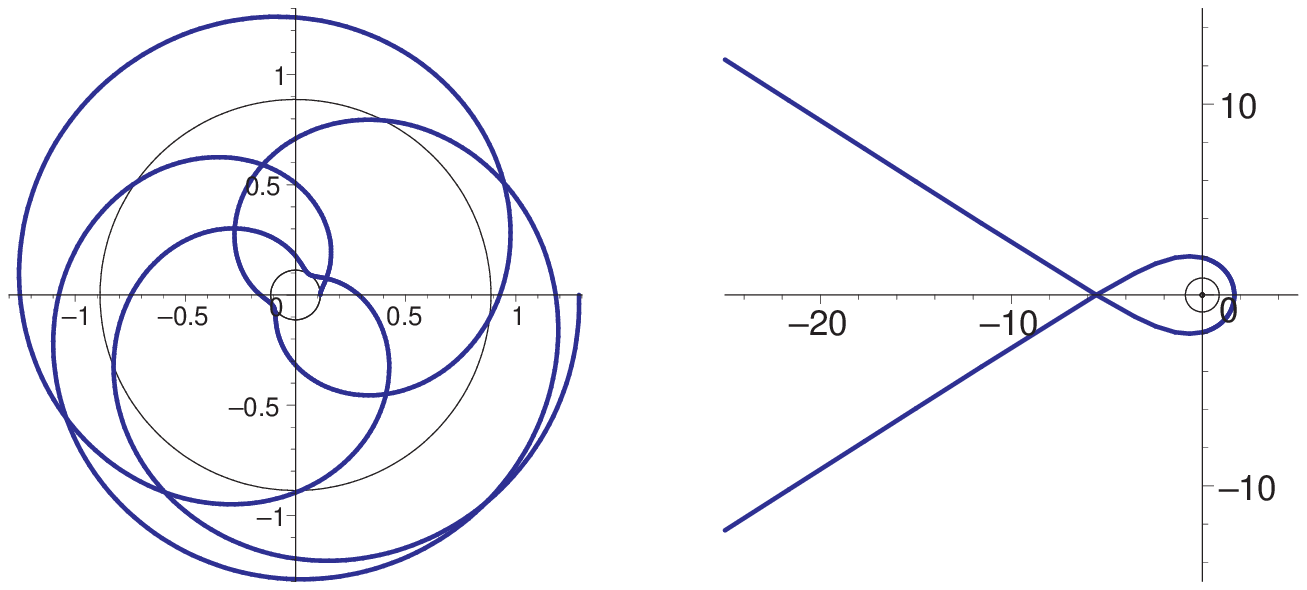}}
\quad
\subfigure[][$\mu=2.1$: two--world escape orbit]{\label{orbRNdS4d:d}%
\includegraphics[width=0.19\textwidth]{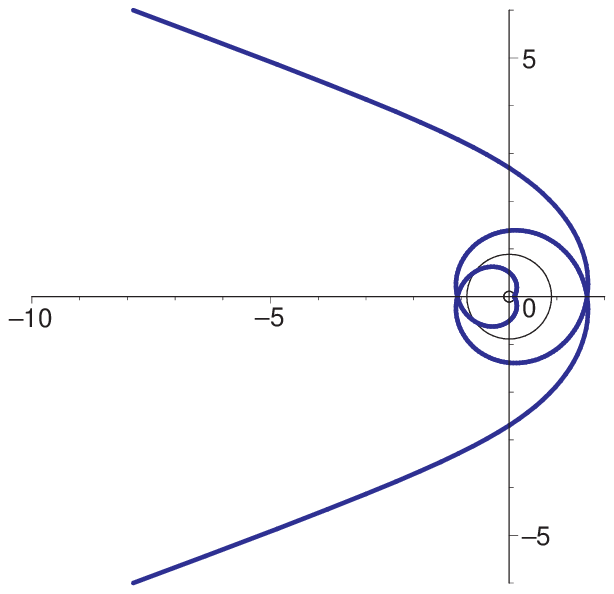}}
\quad
\subfigure[][$\mu=5.0$: two--world escape orbit]{\label{orbRNdS4d:e}%
\includegraphics[width=0.19\textwidth]{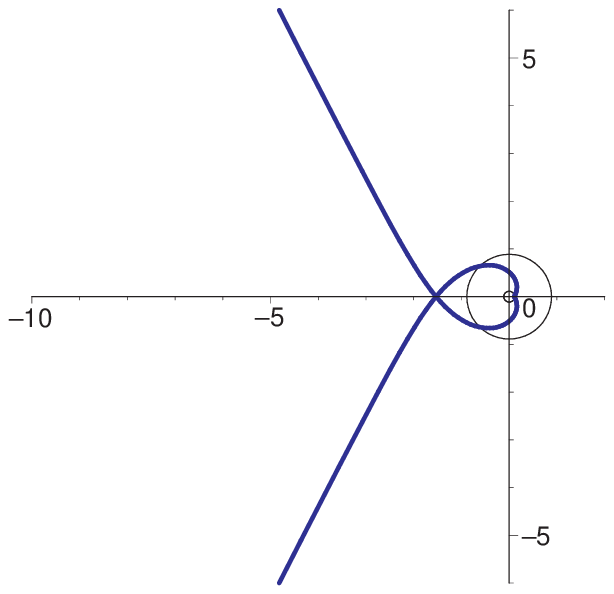}}
\end{center}
\caption{Orbits of a test particle in RN--de Sitter space--time in $4$ dimensions for $\lambda=0.1$ and $\eta=0.1$, $\tilde{\Lambda}=8.7 \cdot 10^{-5}$} \label{RNdSorbd4}
\end{figure}

The substitution $\tilde{r}=-\frac{1}{x}+r_6$, where $r_6$ is a root of $P_6(\tilde{r})$, reduces \eqref{RNdSgeod4d} to $\left(x\frac{dx}{d\varphi}\right)^2 = \frac{ P_5(x)}{3}$ which is of type \eqref{rP_5a} with the general solution given in \eqref{solRN7D}. There are periodic bound orbits presented in the first part of Fig.~\ref{orbRNdS4d:a}--\subref{orbRNdS4d:c} where particles cross both horizons and then enter another universe. Corresponding escape orbits are shown in the second part of Fig.~\ref{orbRNdS4d:a}--\subref{orbRNdS4d:c}. In the second part of the Fig.~\ref{orbRNdS4d:a} the influence of the cosmological constant appears as a barrier for particles coming from infinity. Further two--world escape orbits for particles coming from infinity and crossing the horizons are illustrated in Fig.~\ref{orbRNdS4d:d},\subref{orbRNdS4d:e} for different values of $\mu$.

\begin{figure}[th!]
\begin{center}
\subfigure[][many--world periodic bound orbit]{\label{sub:RNdStwoBOUND_a}%
\includegraphics[width=0.25\textwidth]{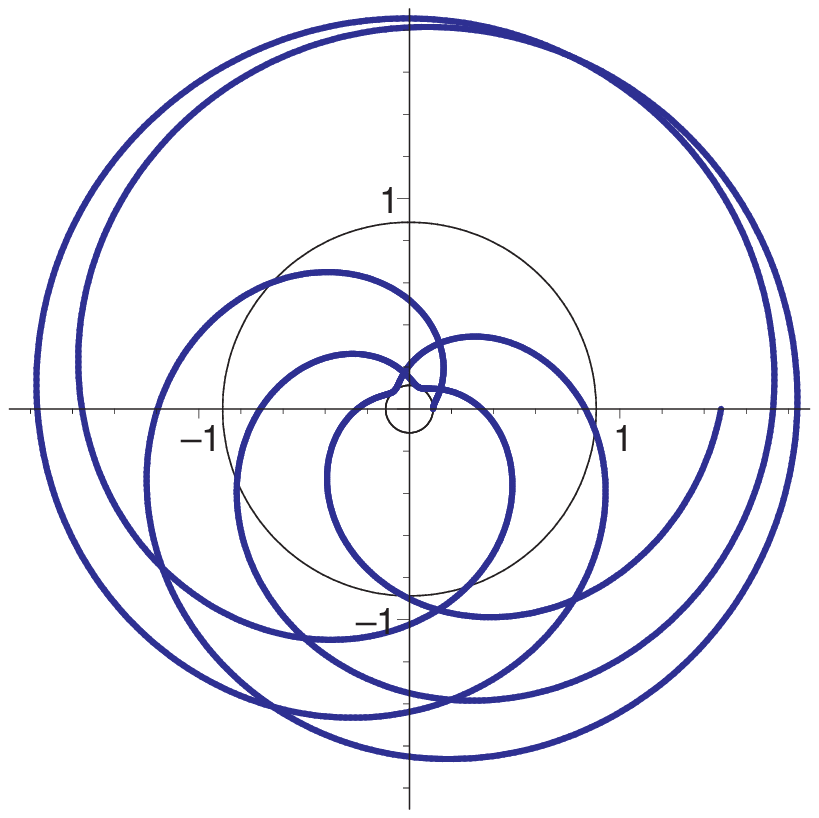}}
\hspace{0.1cm}
\subfigure[][periodic bound orbit]{\label{sub:RNdStwoBOUND_b}%
\includegraphics[width=0.25\textwidth]{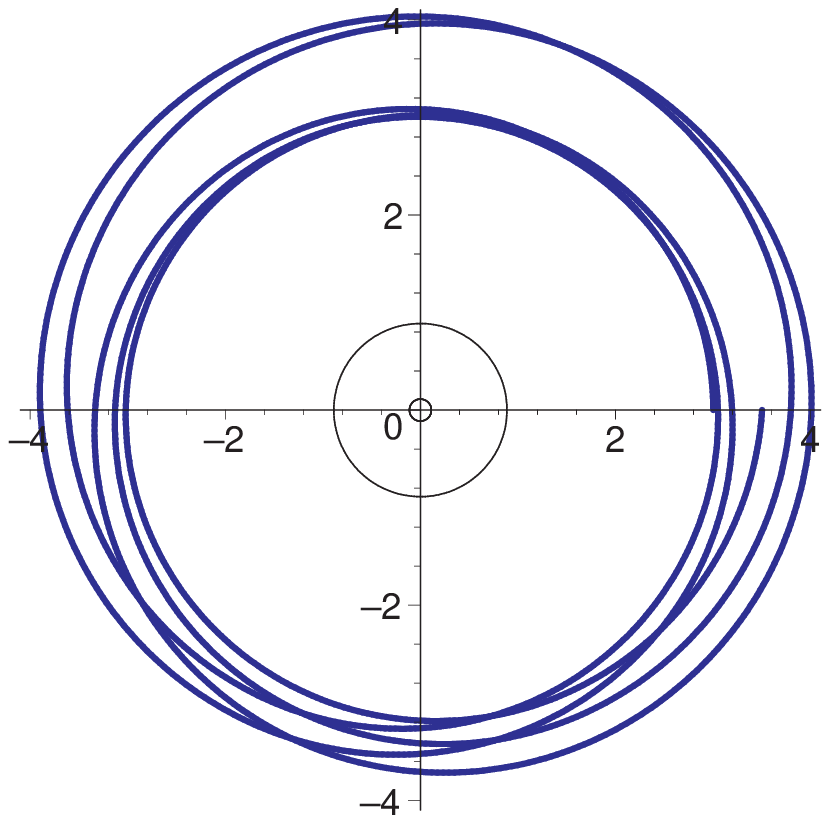}}
\hspace{0.1cm}
\subfigure[][escape orbit]{\label{sub:RNdStwoBOUND_c}%
\includegraphics[width=0.25\textwidth]{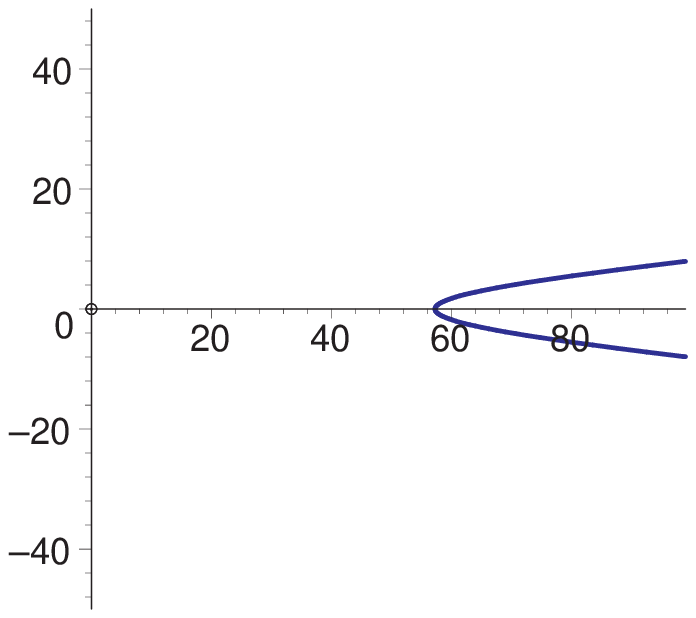}}
\end{center}
\caption{Three geodesics in RN--de Sitter space--time in $4$ dimensions for the parameters  $\lambda=0.3576404$, $\mu=0.888$, $\eta=0.1$, and $\tilde{\Lambda}=8.7 \cdot 10^{-5}$. \label{RNdStwoBOUND}}
\end{figure}

A particular property of $V_{\rm eff}$ is shown in Fig.~\ref{sub:potRNdS4d_2} for $\lambda=0.3576404$ which corresponds to the black area in Fig.~\ref{sub:lamuRNdS4d}. The figure reveals three regions where a test particle can move, and in two of them orbits are bound. The geodesics for $\mu=0.888$ are shown in Fig.~\ref{RNdStwoBOUND} and exhibit a many--world periodic bound orbit in Fig.~\ref{sub:RNdStwoBOUND_a}, an ordinary periodic bound orbit in Fig.~\ref{sub:RNdStwoBOUND_b}, and an escape orbit in Fig.~\ref{sub:RNdStwoBOUND_c}.

Since in general there are two periodic bound orbits we may obtain two perihelia shifts. Each is given by 
\begin{equation}
\Delta_{\rm perihel}^{(j)} =2\pi - 2 \int_{e_{i}}^{e_{i+1}} \frac{x dx}{\sqrt{P_5(x)/3}} = 2\pi - 4\omega_{2k} \ ,
\end{equation}
where the zeros $e_{i}$ and $e_{i+1}$ of $P_5(x)$ again correspond to the considered range of motion $[r_{\rm min}^{(j)}, r_{\rm max}^{(j)}]$ and the path $a_k$ surrounds the interval $[e_{i}, e_{i+1}]$. In our case $j \in \{1\}$ (only one bound orbit) or $j \in \{1, 2\}$ (two bound orbits). While in the 4--dimensional RN space--time for $\mu < 1$ the two periods related to two periodic bound orbits are the same, the appearance of the cosmological constant removes this degeneracy so that the two orbits possess different periods. 

It is again interesting to investigate naked singularities. In the 4--dimensional RN space--time we have two horizons for $\eta < \frac{1}{4}$ (or $q < M$). For $\eta = \frac{1}{4}$ both horizons become degenerate and for $\eta > \frac{1}{4}$ a naked singularity appears.  
 
The horizons for a 4--dimensional RN--de Sitter space--time are given by the real positive zeros of the equation
\begin{equation}
\label{hornak}
\frac{\tilde{\Lambda}}{3}\tilde{r}^4 - \tilde{r}^2 + \tilde{r} - \eta = 0 \ .
\end{equation}
From Descartes' rule we infer: For negative $\tilde\Lambda$ there are two or no positive zeros. For positive $\tilde\Lambda$ there are three or one positive zeros, one of these is the cosmological horizon. For both signs of $\tilde\Lambda$ two zeros may combine to a degenerate horizon.

\begin{figure}[th!]
\begin{center}
\includegraphics[width=0.25\textwidth]{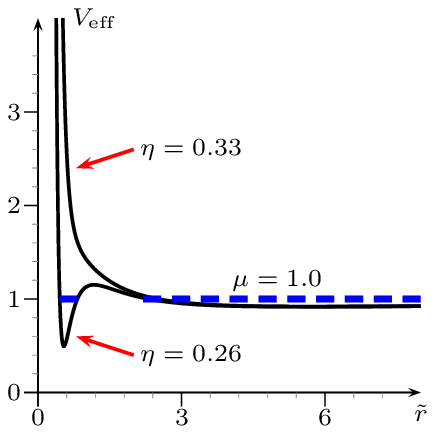}
\end{center}
\caption{Effective potential for naked singularities in RN--de Sitter space--time in $4$ dimensions for $\lambda=0.3$ $\tilde{\Lambda}=8.7 \cdot 10^{-5}$; for orbits see Fig.~\ref{NSinRNdSorbd4}.  \label{NSinRNdS4D}}
\end{figure}

\begin{figure}[th!]
\begin{center}
\subfigure[][$\eta=0.26$, $\mu=1$: periodic bound orbit and escape orbit]{\label{sub:NSinRNdSorbd4_a}%
\includegraphics[width=0.45\textwidth]{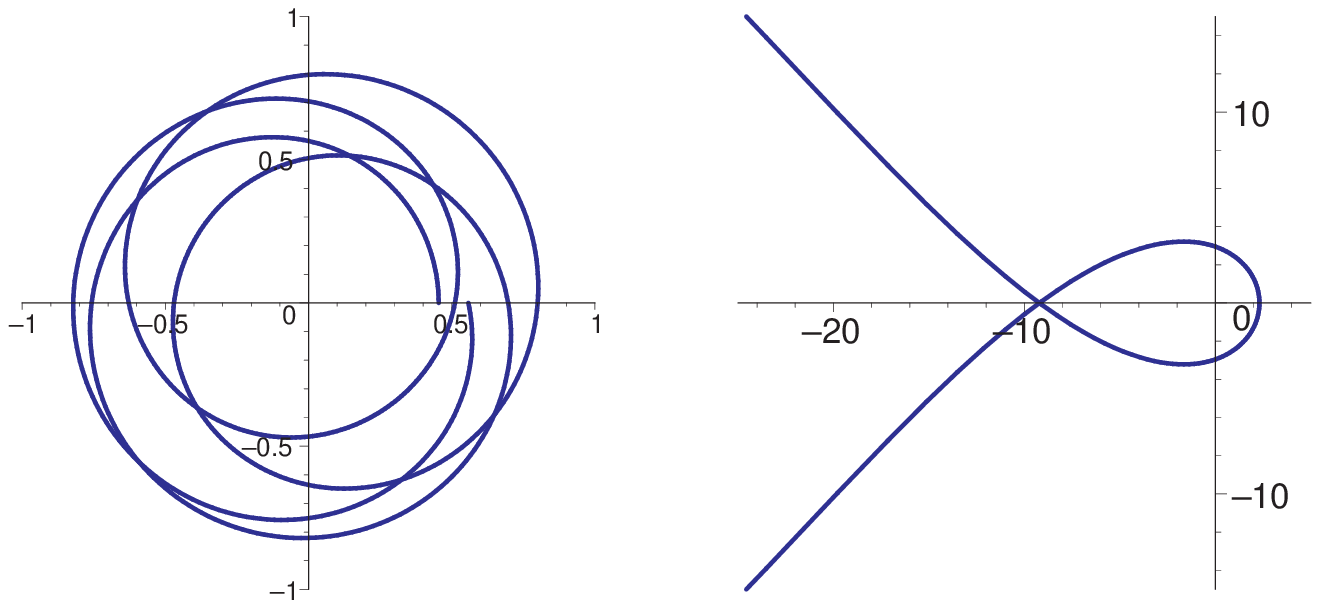}}
\hspace{0.1cm}
\subfigure[][$\eta=0.33$, $\mu=1$: escape orbit]{\label{sub:NSinRNdSorbd4_c}%
\includegraphics[width=0.19\textwidth]{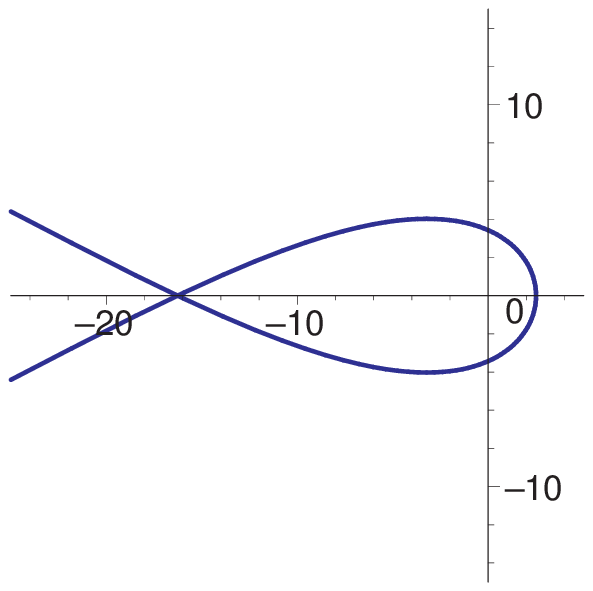}}
\end{center}
\caption{Geodesics in the RN--de Sitter space--time in $4$ dimensions with naked singularity for $\eta=0.26$ and $\eta=0.33$, $\tilde{\Lambda}=8.7 \cdot 10^{-5}$ \label{NSinRNdSorbd4}}
\end{figure}

It is possible to state the algebraic condition on $\eta$ as a function of a given $\tilde\Lambda$ for which a degenerate horizon occurs. In Fig.~\ref{NSinRNdS4D} we show the effective potential~\eqref{Veff4RNdS} for two values of the charge parameter, $\eta=0.26$ and $\eta=0.33$; Fig.~\ref{NSinRNdSorbd4} presents the corresponding orbits. The minimum of $V_{\rm eff}$ is negative for small $\eta$, it vanishes for the critical $\eta$, and it lies above the $\tilde{r}$--axis for larger values of $\eta$. For some larger $\eta$ the minimum disappears. Because of the potential barrier a test particle does not fall into the singularity, but in contrast to the many--world periodic bound orbits discussed above the particle will stay in the same universe (compare Fig.~\ref{sub:NSinRNdSorbd4_a} and e.g. Fig.~\ref{sub:RNdStwoBOUND_a}).

\subsection{Reissner--Nordstr\"om--de Sitter in 7 dimensions}

\begin{figure}[th!]
\begin{center}
\subfigure[][$(\lambda,\mu)$--plot, $\tilde{\Lambda}=8.7 \cdot 10^{-5}$, $\eta=0.4$]{\label{sub:lamuRNdS7d}%
\includegraphics[width=0.3\textwidth]{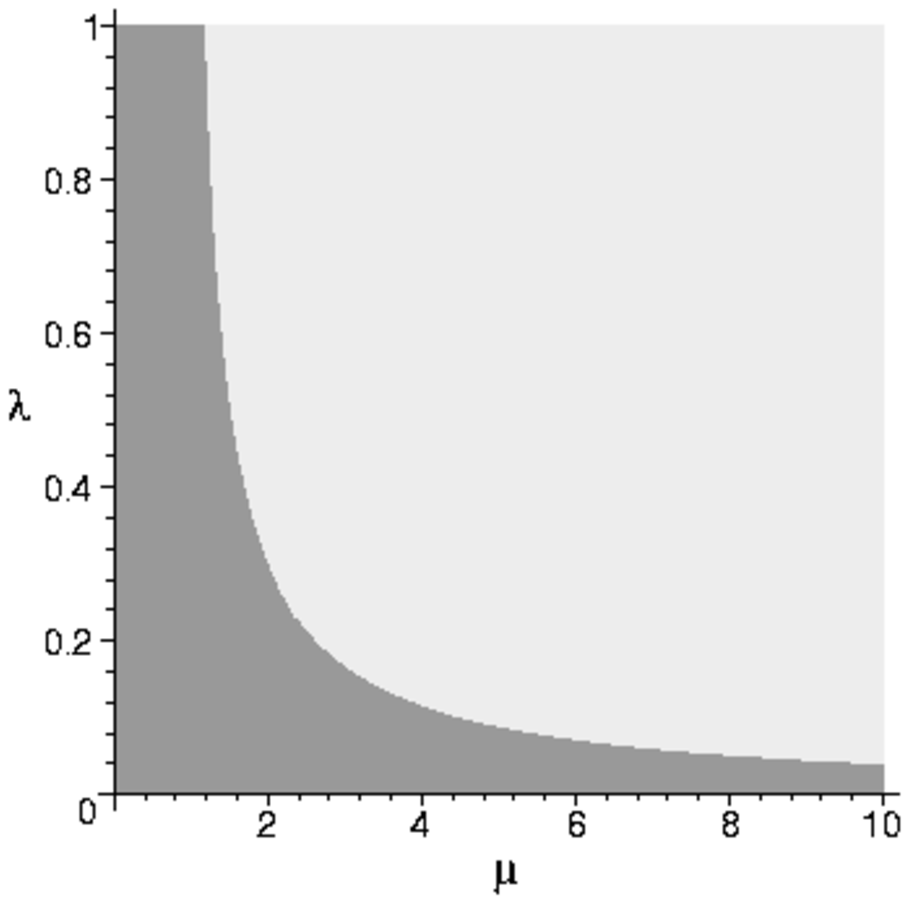}
} \quad %
\subfigure[][Potential for $\lambda=0.05$, $\tilde{\Lambda}=8.7 \cdot 10^{-5}$, $\eta=0.4$]{\label{sub:potRNdS7d}%
\includegraphics[width=0.25\textwidth]{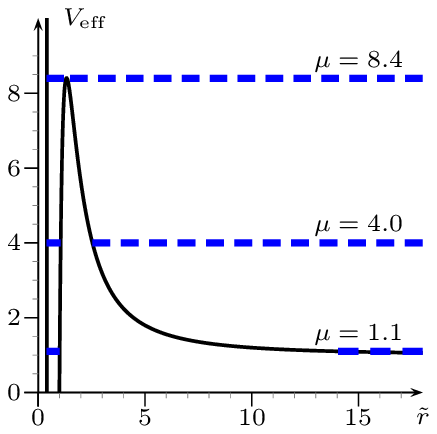}
}
\end{center}
\caption{7--dimensional RN--de Sitter space--time: \subref{sub:lamuRNdS7d} $(\lambda,\mu)$--plot (gray--scales as above). \subref{sub:potRNdS7d} Effective potential for $\eta=0.4$ and $\tilde{\Lambda}=8.7 \cdot 10^{-5}$. For the corresponding orbits see Fig.~\ref{RNdSorbd7}.  \label{lamuRNdS7D}}
\end{figure}

\begin{figure}[th!]
\begin{center}
\subfigure[][$\mu=1.1$: many--world periodic bound orbit and escape orbit]{\includegraphics[width=0.45\textwidth]{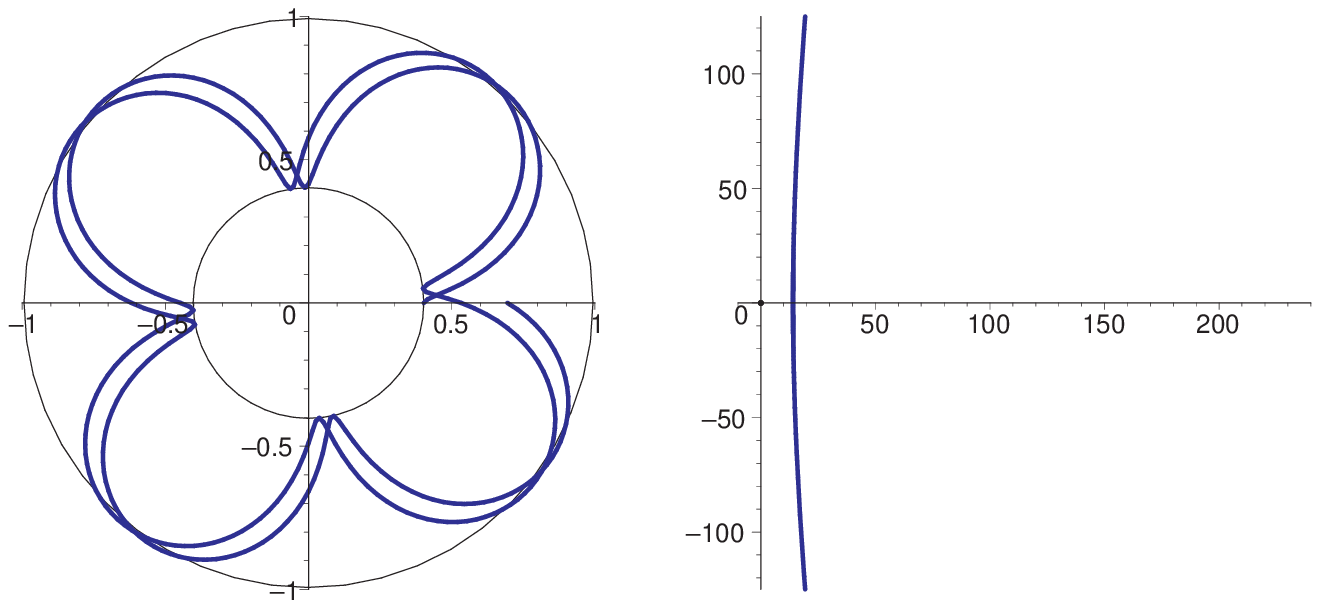}}
\subfigure[][$\mu=4.0$: many--world periodic bound orbit and escape orbit]{\includegraphics[width=0.45\textwidth]{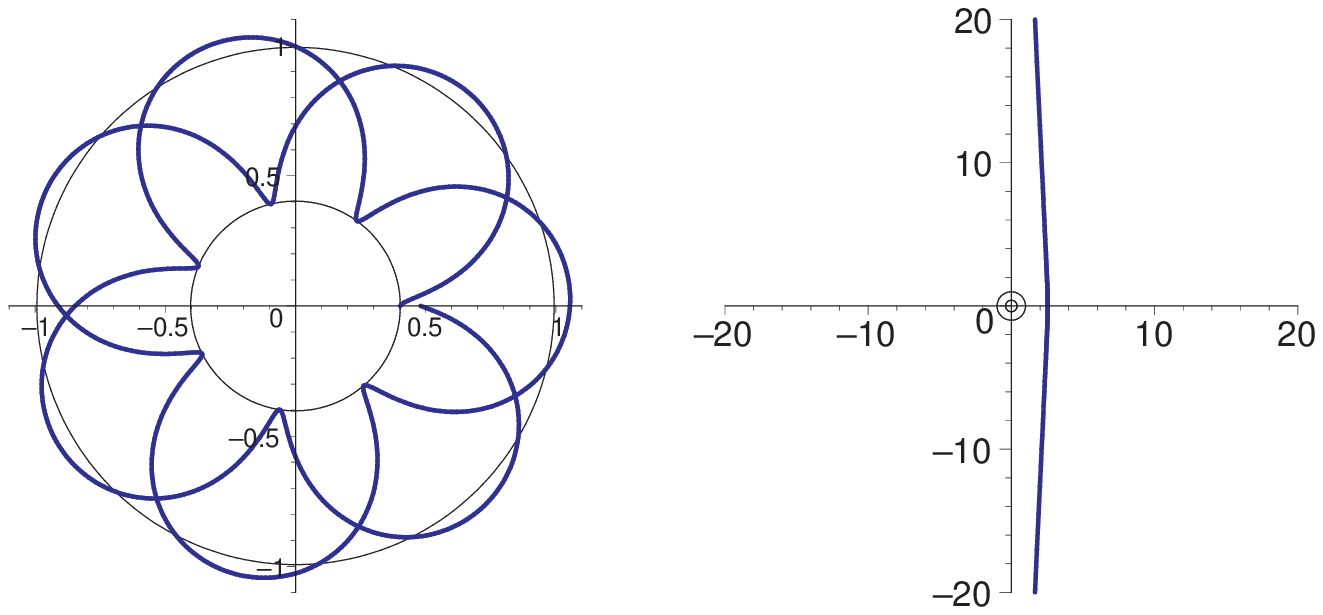}}
\subfigure[][$\mu=8.4$: many--world periodic bound orbit and escape orbit]{\includegraphics[width=0.45\textwidth]{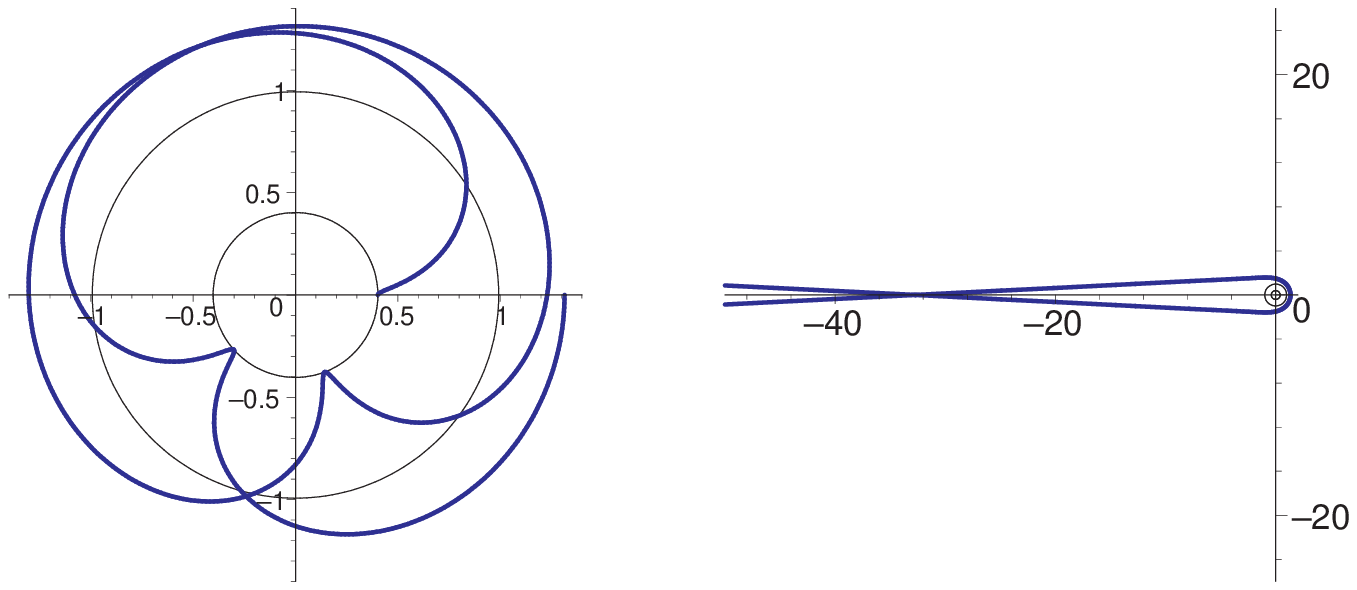}}
\end{center}
\caption{Orbits of a test particle in RN--de Sitter space--time in $7$ dimensions for $\lambda=0.05$ and $\eta=0.4$, and $\tilde{\Lambda}=8.7 \cdot 10^{-5}$.} \label{RNdSorbd7}
\end{figure}

For the 7--dimensional RN--de Sitter space--time equation~\eqref{EOMrphinormu} becomes
\begin{equation}
\label{RNdSgeod7d}
\left(\frac{du}{d\varphi}\right)^2= 4 \Bigg( - \eta^4 u^6 - \eta^4 \lambda u^5+ u^4+ \lambda u^3- u^2+\left( \lambda (\mu -1) + \frac{\tilde{\Lambda}}{15} \right) u+ \frac{\tilde{\Lambda}}{15} \lambda  \Bigg) = 4 P_6(u) \ .
\end{equation}
The number of positive zeros of $P_6(u)$ (or $P_{12}(\tilde{r})$) is drawn in Fig.~\ref{sub:lamuRNdS7d} and the effective potential 
\begin{equation}
\label{Veff7RNdS}
V_{\rm eff}= \left(1 - \frac{1}{\tilde{r}^4} - \frac{\tilde{\Lambda}}{15}\tilde{r}^2 +  \frac{\eta^4}{\tilde{r}^8} \right) \left(1 + \frac{1}{\lambda\tilde{r}^2} \right) \,
\end{equation}
for $\lambda=0.05$, $\eta=0.4$ and $\tilde{\Lambda}=8.7\cdot 10^{-5}$ in Fig.~\ref{sub:potRNdS7d}. The dark gray area of 3 real positive roots corresponds to many--world periodic bound orbits and escape orbits shown in Fig.~\ref{RNdSorbd7}, and the gray area to two--world escape orbits. 

Introduction of a new variable $x$ such that $u=-\frac{1}{x}+u_6$, where $u_6$ is the root of $P_6(u)$, reduces ~\eqref{RNdSgeod7d} to $\left(x\frac{dx}{d\varphi}\right)^2= 4 P_5(x)$ which is of type~\eqref{rP_5a} with the solution given in \eqref{solRN7D}.

For a periodic bound orbit in the interval between points $e_i$ and $e_{i+1}$ and the path $a_k$ surrounding this interval the perihelion shift $\Delta_{\rm perihel}$ is defined as
\begin{equation}
\Delta_{\rm perihel}=2\pi - 2 \int_{e_i}^{e_{i+1}} \frac{x dx}{\sqrt{4 P_5(x)}} = 2\pi - 4 \omega_{2 k} \ .
\end{equation}

\section{Light rays}

For light rays $\epsilon = 0$ and the geodesic equation \eqref{EOMrphinorm} reduces to 
\begin{equation}
\left(\frac{d\tilde{r}}{d\varphi}\right)^2 = \tilde{r}^4 \left(\lambda \mu + \frac{2 \tilde\Lambda}{(d-1)(d-2)} - \left(1 - \frac{1}{\tilde{r}^{d-3}} + \frac{\eta^{d-3}}{\tilde{r}^{2(d-3)}}\right) \frac{1}{\tilde{r}^2}\right) \, . \label{EOMrphinormLight}
\end{equation}
Here the cosmological constant only contributes to the energy parameter, it does not change the mathematical structure of the differential equation. The only remaining parameter of relevance is $\eta$. For $\eta = 0$ the equation has the structure $\left(\frac{d\tilde{r}}{d\varphi}\right)^2 = P_{d-1}(\tilde{r})/\tilde{r}^{d - 5}$ and for $\eta \neq 0$ we have $\left(\frac{d\tilde{r}}{d\varphi}\right)^2 = P_{2(d-2)}(\tilde{r})/\tilde{r}^{2(d - 4)}$. For a space--time with odd dimension $d$ we get for \eqref{EOMrphinormu}
\begin{equation}
\left(\frac{du}{d\varphi}\right)^2 = 4 u \left(- \eta^{d-3} u^{d-2} + u^{\frac{1}{2} (d-1)} - u + \lambda\mu + \frac{2 \tilde\Lambda}{(d-1)(d-2)}\right) \, . \label{EOMuphiLight}
\end{equation}

The effective potential for light is
\begin{equation}
V_{\rm eff} = f(r) \frac{L^2}{r^2} = - \frac{2 \tilde\Lambda}{\lambda (d - 1) (d-2)} + \frac{1}{\lambda \tilde{r}^2} - \frac{1}{\lambda \tilde{r}^{d-1}} + \frac{\eta^{2(d-3)}}{\lambda \tilde{r}^{2(d-2)}} \, .
\end{equation}
It is clear that the cosmological constant does not play any role in the motion of light rays. 
For higher dimensions the attractive $-1/\tilde{r}^{d-1}$ becomes more pronounced. Also the repulsive part related to the charge becomes stronger for higher dimensions and also for larger charges. 

In the case of Schwarzschild and S(a)dS space--times the equations of motion are
\begin{equation}
\left(\frac{d\tilde{r}}{d\varphi}\right)^2 = \frac{1}{\tilde{r}^{d - 5}} \left(\left(\lambda \mu + \frac{2 \tilde\Lambda}{(d-1)(d-2)}\right) \tilde{r}^{d - 1} - \tilde{r}^{d - 3} + 1\right) \ ,
\end{equation}
here are two sign changes in the polynomial on the right--hand--side indicating a terminating bound and an escape orbit, or a terminating escape orbit. This can by verified by the form of the effective potential. The equation of motion can be analytically solved with elliptic functions for 4, 5, 7 and with hyperelliptic functions for 6, 9 and 11 dimensions. No analytic solution is known for 8, 10, 12 and all higher dimensions. 

In a RN or RN(a)dS space--time the equation of motion has the form
\begin{equation}
\left(\frac{d\tilde{r}}{d\varphi}\right)^2 = \frac{1}{\tilde{r}^{2(d-4)}} \left(\left(\lambda \mu + \frac{2 \tilde\Lambda}{(d-1)(d-2)}\right) \tilde{r}^{2(d-2)} - \tilde{r}^{2(d-3)} + \tilde{r}^{d-3} - \eta^{d-3}\right) \, .
\end{equation}
If the first term is positive then Descartes' rule states that there are three or one real positive zeros of the polynomial implying the existence of many--world periodic bound and escape orbits or two--world escape orbits. That means light can disappear into another universe or appear from another universe. For a large negative $\tilde\Lambda$ escape orbits no longer exist. This can also be seen from the corresponding effective potential. Using \eqref{EOMuphiLight} the equation of motion can be solved for 5 and 7 dimensions, in 5 dimensions by elliptic functions, in 7 dimensions with hyperelliptic functions as shown above. 

\section{Radial motion}

For radial motion, i.e., for $L = 0$, the geodesic equation~\eqref{drds} for point particles reduces to 
\begin{equation}
\left(\frac{d\tilde{r}}{ds}\right)^2 = \mu - 1 + \frac{1}{\tilde{r}^{d-3}} + \frac{2 \tilde\Lambda\tilde{r}^2}{(d-1)(d-2)} - \frac{\eta^{d-3}}{\tilde{r}^{2(d-3)}} \, .
\end{equation}
This equation can be integrated along the lines presented above for non--radial motion. The effective potential for this radial motion is 
\begin{equation}
V_{\rm eff} = 1 - \frac{1}{\tilde{r}^{d-3}} - \frac{2 \tilde\Lambda\tilde{r}^2}{(d-1)(d-2)} + \frac{\eta^{d-3}}{\tilde{r}^{2(d-3)}}
\end{equation}
Of interest are the equilibrium positions given by the vanishing of the first derivative of the effective potential
\begin{equation}
\label{radmot}
0 = (d - 3) \tilde{r}^{d - 3} - \frac{4 \tilde\Lambda}{(d-1)(d-2)} \tilde{r}^{2 (d - 2)} - 2 (d-3) \eta^{d-3} \, .
\end{equation}
If $\eta = 0$ then there is a solution only for $\Lambda > 0$ which, however, corresponds to a maximum and, thus, gives no stable position. This equation has stable solutions only if $\eta \neq 0$, independent of the dimension. For simplicity, we now choose $\Lambda=0$. Then \eqref{radmot} has the solution $\tilde{r}_{0}= 2^{\frac{1}{d-3}} \eta$ with the effective potential $V_{\rm eff}(\tilde{r}_{0})=1-\frac{1}{4}\eta^{3-d}$. Therefore, in any dimension for $\eta\ge 4^{\frac{1}{3-d}}$ a charged solution allows particles to stay at rest at a stable position at a finite value of the radial coordinate.

\section{Conclusion and outlook}\label{Sec:conclusion}

A given gravitational field can only be analyzed and interpreted through the exploration of the geodesics of particles and light rays. In this paper we have discussed the motion of test particles and light rays in higher dimensional space--times of spherically symmetric gravitational sources endowed with mass and electric charge and a cosmological constant. We discussed the general structure of the resulting orbits. One result of this analysis is that only in 4--dimensional Schwarzschild and Schwarzschild--de Sitter space--times it is possible to have stable periodic orbits which do not cross horizons. In Reissner--Nordstr\"om space--times of any dimension we have stable orbits which, however, periodically disappear into other universes. This raises questions about the operational realization and interpretation of notions like the perihelion shift and scattering angles. Furthermore, due to the structure of the angular momentum barrier and the cosmological force only in 4 and 5 dimension we have a rich variety of orbits. For 6 and higher dimensions we always encounter the same two types of orbits. 

We analytically integrated the equations of motion for test particles and light rays in Schwarzschild and Schwarzschild--de Sitter space--times in $9$ and $11$ dimensions, in Reissner--Nordstr\"om space-times in $7$ and Reissner--Nordstr\"om--de Sitter space--times in $4$ and $7$ dimensions by applying a method based on a solution of Jacobi inversion problem restricted to the set of zeros of a theta function and described in~\cite{HackmannLaemmerzahl08_PRL, HackmannLaemmerzahl08_PRD}. The explicit integration was possible because in these cases the underlying polynomial appearing in the equation of motion is at most of order 6. 

The next step would be to find the solutions of geodesic equations with an underlying polynomial of 7th and higher order (cases indicated with ''--`` in Table~\ref{tabelle}). In these cases we have to enlarge the number of variables to be three or more. It is not clear currently how to constrain the Abel mapping between 3 or higher dimensional spaces in order to reduce the number of variables.

Meanwhile, it is of special importance to find analytical solutions of the equations of motion. Because of the, in principle, arbitrary high accuracy of analytical solutions one can test with these solutions  numerical codes, e.g., for the dynamics of binary systems in EMRIs (Extreme-Mass-Ratio Inspirals). Another application is that the offered solution can serve as starting point for an advanced approximation scheme in more complicated situations appearing in stellar, planetary, comet, asteroid or satellite dynamics.

\begin{acknowledgements}
We are deeply grateful to W. Fischer and P. Richter for valuable discussions. V.K. thanks the German Academic Exchange Service DAAD and C.L. the German Aerospace Center DLR for financial support.
\end{acknowledgements}

\labelsep20pt

\end{document}